\documentclass[twocolumn]{aastex6}

\newcommand{\sunrise}{\textsc{Sunrise}}

\shorttitle{The second flight of \sunrise{}}

\begin{document}

\title{The second flight of the \sunrise{} balloon-borne solar observatory:
overview of instrument updates, the flight, the data and first results}

\author{\textsc{
S.~K.~Solanki,$^{1,2}$
T.~L.~Riethm\"uller,$^{1}$
P.~Barthol,$^{1}$
S.~Danilovic,$^{1}$
W.~Deutsch,$^{1}$
H.-P.~Doerr,$^{1,6}$
A.~Feller,$^{1}$
A.~Gandorfer,$^{1}$
D.~Germerott,$^{1}$
L.~Gizon,$^{1,3}$
B.~Grauf,$^{1}$
K.~Heerlein,$^{1}$
J.~Hirzberger,$^{1}$
M.~Kolleck,$^{1}$
A.~Lagg,$^{1}$
R.~Meller,$^{1}$
G.~Tomasch,$^{1}$
M.~van~Noort,$^{1}$
J.~Blanco~Rodr\'{\i}guez,$^{4}$
J.~L.~Gasent Blesa,$^{4}$
M.~Balaguer Jim\'enez,$^{5}$
J.~C.~Del~Toro Iniesta,$^{5}$
A.~C.~L\'opez Jim\'enez,$^{5}$
D.~Orozco~Su\'arez,$^{5}$
T.~Berkefeld,$^{6}$
C.~Halbgewachs,$^{6}$
W.~Schmidt,$^{6}$
A.~\'Alvarez-Herrero,$^{7}$
L.~Sabau-Graziati,$^{7}$
I.~P\'erez Grande,$^{8}$
V.~Mart\'{\i}nez Pillet,$^{9}$
G.~Card,$^{10}$
R.~Centeno,$^{10}$
M.~Kn\"olker,$^{10}$
\& A.~Lecinski$^{10}$
}}

\affil{
$^{1}$Max Planck Institute for Solar System Research, Justus-von-Liebig-Weg 3,
      37077 G\"ottingen, Germany; solanki@mps.mpg.de\\
$^{2}$School of Space Research, Kyung Hee University, Yongin, Gyeonggi, 446-701,
      Republic of Korea\\
$^{3}$Institut f\"ur Astrophysik, Georg-August-Universit\"at G\"ottingen,
      Friedrich-Hund-Platz 1, 37077 G\"ottingen, Germany\\
$^{4}$Grupo de Astronom\'{\i}a y Ciencias del Espacio, Universidad de Valencia,
      46980 Paterna, Valencia, Spain\\
$^{5}$Instituto de Astrof\'{\i}sica de Andaluc\'{\i}a (CSIC),
      Apartado de Correos 3004, 18080 Granada, Spain\\
$^{6}$Kiepenheuer-Institut f\"ur Sonnenphysik, Sch\"oneckstr. 6, 79104 Freiburg, Germany\\
$^{7}$Instituto Nacional de T\'ecnica Aeroespacial, Carretera de Ajalvir, km 4,
      28850 Torrej\'on de Ardoz, Spain\\
$^{8}$Universidad Polit\'ecnica de Madrid, IDR/UPM, Plaza Cardenal Cisneros 3,
      28040 Madrid, Spain\\
$^{9}$National Solar Observatory, 3665 Discovery Drive, Boulder, CO 80303, USA\\
      $^{10}$High Altitude Observatory, National Center for Atmospheric Research,
P.O. Box 3000, Boulder, CO 80307-3000, USA\\
}

%

\begin{abstract}
The \sunrise{} balloon-borne solar observatory, consisting of a 1~m aperture
telescope that provided a stabilized image to a UV filter imager and an
imaging vector polarimeter, carried out its second science flight in June 2013. It
provided observations of parts of active regions at  high spatial resolution,
including the first high-resolution images in the Mg~{\sc ii}~k line. The
obtained data are of very high quality, with the best UV images reaching the
diffraction limit of the telescope at 3000~\AA\ after Multi-Frame Blind
Deconvolution reconstruction accounting for phase-diversity information.
Here a brief update is given of the instruments and the data reduction
techniques, which includes an inversion of the polarimetric data. Mainly those
aspects that evolved compared with the first flight are described.
A tabular overview of the observations is given. In addition, an example
time series of a part of the emerging active region NOAA AR~11768
observed relatively close to disk centre is described and discussed in some
detail. The observations cover the pores in the trailing polarity of the
active region, as well as the polarity inversion line where flux emergence
was ongoing and a small flare-like brightening occurred in the course of the
time series. The pores are found to
contain magnetic field strengths ranging up to 2500~G and, while large pores
are clearly darker and cooler than the quiet Sun in all
layers of the photosphere, the temperature and brightness of small pores
approach or even exceed those of the quiet Sun in the upper
photosphere.
\end{abstract}

\keywords{Sun: chromosphere -- Sun: faculae, plages -- Sun: sunspots, pores -- Sun: photosphere
-- techniques: photometric -- techniques: polarimetric -- techniques: spectroscopic}

\section{Introduction} \label{sec:intro}

Probing the Sun at high resolution has time and again revealed new phenomena
not previously seen. Examples are the discovery of umbral dots by \citet{Beckers1968},
facular bright points by
\citet{Mehltretter1974}, dark cores in penumbral
filaments by \citet{Scharmer2002}, lateral downflows in penumbral filaments
by \citet{Joshi2011a, Joshi2011b, Scharmer2011}, or the ultrafine
loops reported by \cite{Ji2012}.

In spite of the significant advances made in ground-based observations, it
remains challenging to accurately take into account or fully remove the effects
of atmospheric seeing from observational data. In addition, high-resolution
studies from the ground are generally limited to wavelengths with
high photon flux and usually to short periods of stable seeing. Thus, in
spite of having an aperture smaller than that of the largest ground-based
telescopes, the space-based Solar Optical Telescope onboard Hinode
\citep{Tsuneta2008} has resulted in many advances. These include the
discovery of penumbral microjets \citep{Katsukawa2007}, waves carrying
copious amounts of energy along spicules \citep{DePontieu2007}, ubiquitous
linear-polarisation signals in the quiet Sun \citep{Orozco2007, Lites2008, Lagg2016},
or very fast downflows at the ends of particular penumbral filaments
\citep{vanNoort2013}. Hence, an even larger solar telescope located above
the bulk of the Earth's atmosphere has an extensive discovery space.

The largest solar telescope \citep[along with the Soviet stratospheric solar
observatory;][]{Krat1974} to have reached
the near-space conditions of the stratosphere is \sunrise{}, which had a very
successful first flight in June 2009 \citep[for an overview of earlier balloon-borne
solar telescopes and results see][]{Solanki2010}.
The data obtained during that first flight of \sunrise{} has led to the
following discoveries and insights, among many other results:

\begin{itemize}
\item First ever spatially resolved images of small-scale intense magnetic flux
concentrations in the quiet Sun show that semi-empirical flux tube models provide a
reasonable description of such structures \citep{Lagg2010}.

\item First ever brightness measurements of flux concentrations in the UV at
312 nm, 300 nm and 214 nm reveal very high intensities, up to a factor of 5 above
the mean quiet-Sun brightness at 214 nm \citep{Riethmueller2010}.

\item First ever measurements of the RMS intensity contrast of granulation in
the UV show high values of up to 30\%, consistent with numerical simulations.
These values provide a direct measure of the efficiency of convective energy
transport by granulation \citep{Hirzberger2010}.

\item The most sensitive high-resolution time sequences of maps of the vector
magnetic field ever obtained reveal abundant, short-lived and highly dynamic
small-scale horizontal fields \citep{Danilovic2010}.

\item Ubiquitous small-scale whirl flows are found which drag small-scale
magnetic field structures into their centres \citep{Bonet2010}.


\item Magnetic field extrapolations from \sunrise{}/IMaX data indicate that
most magnetic loops in the
quiet Sun remain within the photosphere. Only a small fraction reaches the
chromosphere. Most of these higher-lying loops are anchored (at least in one
foot point) in the strong-field elements of the network \citep{Wiegelmann2010}.

\item Discovery of large amplitude acoustic waves in the
quiet solar atmosphere. Such waves were missed in the past,
since they are spatially strongly localized and their photospheric sources move
significantly within a short time \citep{BelloGonzalez2010}.

\item First detection of horizontally oriented vortex tubes in solar convective
features. Such vortex tubes were found to be rather common in solar granules
\citep{Steiner2010}.

\item Discovery that the internetwork magnetic elements continuously move back
and forth between a state of weak and strong magnetic field
\citep{MartinezGonzalez2011, Requerey2014}.

\item First determination of the inclination of magnetic elements directly from
their position in images sampling different heights. The results reveal that the
magnetic elements are nearly vertical, in contrast to inversion techniques that
suggested that they were close to horizontal \citep[but in this case were strongly
affected by noise;][]{Jafarzadeh2014b}.

\item First detection of localized, strongly wavelength-shifted polarization
signals in the quiet
Sun that are interpreted as supersonic upflows caused by magnetic reconnection
of emerging small-scale loops with pre-existing fields \citep{Borrero2010},
confirmed by Hinode observations \citep{MartinezPillet2011b}.

\item Discovery that 85 \%\ of the internetwork magnetic fields stronger than
100 G are concentrated in mesogranular lanes, although there is no particular
mesogranular scale \citep{Yelles2011}.
\end{itemize}

The data taken during the first flight of the \sunrise{} observatory were
limited to the quiet Sun, as the Sun was exceedingly quiet for
the whole duration of the flight. To be able to probe an active region with
the unique capabilities of
the \sunrise{} observatory, a reflight of the largely (but not completely)
unchanged instrumentation was carried out in June 2013. In this paper we describe
the updates of the instruments and of the data reduction. We also provide an
overview of the obtained data and of the first results
obtained from this second flight of \sunrise{}. A set of results
from this second flight are described in the publications
that are part of this special issue (which also contains some papers making
use of data from the first \sunrise{} science flight).

In the following we shall often refer to the instrumentation as flown during the
first flight in 2009 and the corresponding mission as \sunrise{}~I, while the updated
instrumentation flown in 2013 and that flight are referred to as \sunrise{}~II.

The paper is structured as follows. In Sect.~\ref{sec:instrum} we provide a
summary of the updates to the instrumentation, concentrating mainly on changes
made relative to the first flight in 2009. In Sect.~\ref{sec:mission}, the 2013
flight is described and an overview of the obtained data is given in
Sect.~\ref{sec:overview}. The updated data reduction procedures are described in
Sect.~\ref{sec:reduction}, again focussing on the parts that changed relative to
the first flight. Some snapshots of the reduced data are discussed in
Sect.~\ref{sec:results}. Finally, conclusions and an outlook on future plans for
\sunrise{} are presented in Sect.~\ref{sec:conclusions}.

\section{Instrumentation update} \label{sec:instrum}

\sunrise{}~II is composed of a Gregory telescope with a main mirror of 1~m
diameter and two post-focus instruments located above it, together with the
Image Stabilization and Light Distribution unit (ISLiD) and the
Correlating Wavefront Sensor (CWS). The telescope and
instruments are mounted in a gondola hanging under a zero-pressure balloon.
The system is stabilized by the gondola and a tip-tilt mirror located in
ISLiD that is controlled by
the CWS. The post-focus instruments are the \sunrise{} Filter Imager (SuFI), that
observes in 5 wavelength channels between roughly 2000 and 4000~\AA, and a filter
magnetograph, the Imaging Magnetograph eXperiment (IMaX), observing in the
Fe~{\sc i} 5250.2~\AA\ spectral line, a Zeeman triplet with a
Land\'e factor of $g=3$. The complete instrumentation is described in brief by
\cite{Solanki2010} and in greater detail by \cite{Barthol2011}. Further
information on ISLiD and on SuFI is given by
\cite{Gandorfer2011}, IMaX is described by \cite{MartinezPillet2011a} and more
information on the CWS is provided by
\cite{Berkefeld2011}.

The major components of \sunrise{}~I, such as the gondola, the telescope and the
post-focus instrumentation suffered
only minor damage during the landing after the first science flight in 2009 and
could be reworked and prepared for a reflight with comparatively little effort.
Below we describe the main
changes to the hardware that were made in preparation for the second flight.

\subsection{Gondola, telemetry and telescope}

{\it Gondola:}
Structural protective elements on the gondola, such as the crush pad assembly below
the core gondola frame or the front and rear roll cages needed to be replaced,
as they were designed to deform and thus to take most of the impact energy to
protect the rest of the payload. In addition to such refurbishment,
modifications were implemented to the gondola, as described below.

The average power consumption during the first flight was much lower than
previously estimated, so that the
number of solar cells could be reduced to only 6 panels with 80
SunPower A-300 cells each (from originally 10 panels in 2009). The panels were
mechanically re-arranged, reducing the overall width of the instrument. This
modification allowed a pre-installation of the solar panels in the integration
hall before roll-out, saving precious time on the day of the launch.
Additional benefits are reduced mass and a lower aerodynamic cross section.

The mounting of the electronics racks, carrying all the instrument computers,
pointing system computer and power distribution units, was modified for the 2013
flight. Instead of having a $20^\circ$ tilt towards cold sky as in the previous
configuration, the racks were now mounted vertically and were directly
attached to the gondola side trusses, providing higher overall stiffness and
reduced mass. The reduction in thermal efficiency of this configuration was seen to be
acceptable, as results from the 2009 flight indicated that the size of the radiating
surfaces are sufficiently large to dump all dissipated heat at moderate temperatures
even in this slightly less efficient configuration.

As part of the on-ground testing before flight, a modal survey test was performed
to determine the first natural eigenmodes of the complete instrument.
A set of externally mounted accelerometers recorded the impulse and frequency
response to an impact on the gondola structure. It was verified that the 10 Hz
oscillation observed in the pointing data of the 2009 science flight corresponds
to the first torsional bending mode of the gondola core framework. Modifications
of software filters were then implemented into the pointing system control loops,
preventing excitation of this mode during the 2013 flight. Pointing system
housekeeping data taken during the 2013 flight clearly demonstrated their efficiency,
as almost no 10 Hz oscillations were present during the time spent at float altitude
and in particular not when pointing at the Sun.

{\it Telemetry:}
The 2009 flight operations and commissioning were hampered by the early loss
of the E-Link high-speed telemetry provided by ESRANGE,\footnote{The ESRANGE Space
Center (67.89$^\circ$N, 21.10$^\circ$E) near Kiruna, Northern Sweden is a European research
center and launch facility for sounding rockets and balloons that is managed by the
Swedish Space Corporation. ESRANGE is financed by the Esrange And{\o}ya Special
Project (EASP) within ESA (European Space Agency). The member states of
ESA/EASP today are France, Germany, Switzerland, Norway and Sweden.} which ceased operation
after only a few hours when switching from one ground station to another. For
the rest of the 2009 mission commanding and health status information was
transmitted through a 6\,kbit/s TDRSS link, where TDRSS is the Tracking and
Data Relay Satellite System, a NASA network of communication satellites on
geosynchronous orbits and associated ground stations used for space-to-ground
communication. This was found to be clearly insufficient.

In preparation for the 2013 science flight the \sunrise{} team helped to qualify the Iridium
Pilot/OpenPort system for balloon flights. A dedicated test set-up was flown
on Sept. 28 and Oct. 07, 2011 from Ft. Sumner on balloons of the Columbia
Scientific Ballooning Facility (CSBF) of NASA for several
hours each, to verify system integrity during launch, ascent and operational
conditions. The bandwidth transmitted to ground was more than ten times higher
than with the TDRSS Omni link.
\sunrise{}~II was one of the first scientific missions to be supported with this
new telemetry system. The antenna system on top of the gondola was re-arranged
to provide the required 6 ft clearance to the dome-shaped Iridium transmitter.
\sunrise{}~II
reached an average data rate of 100\,kbit/s, and downlink of science data continued
even after the payload landed, although the antenna was partly buried
in Canadian soil.

{\it Telescope and structure:}
The telescope and instrumentation needed a thorough cleaning and refurbishment.
All mechanisms were disassembled, inspected, cleaned, lubricated, re-assembled
and requalified. Motor and gears of the telescope aperture mechanism were
replaced and the mechanism was improved to provide a more reliable detection of
the open or closed position.
Key elements of the carbon-fiber based structure were inspected, strength
tested and found to be still fully intact. The thermal subsystem consisting
of the heat rejection wedge and its radiators needed to be refurbished. New
second surface mirrors were attached.
All telescope mirror coatings were stripped, the mirrors cleaned, their optical
quality interferometrically verified and recoated. The secondary mirror needed
to be replaced completely, as the Zerodur mirror substrate was damaged during
the landing and recovery after the 2009 flight. The mirror was refabricated
by the Lytkarino Optical Glass Factory (LZOS) in Moscow, with a
performance identical to the original one.

The telescope alignment proved to be a challenge, because a sufficiently large
reference flat mirror provided by SAGEM for the 2009 flight was not available this time.
Since on the launch site a misalignment of the telescope was detected
(the telescope had been tested before departure to Kiruna without a 1m reference
flat, making the performance assessment ambiguous), a readjustment of the M2
position was necessary.

\begin{figure}
\figurenum{1}
\gridline{\fig{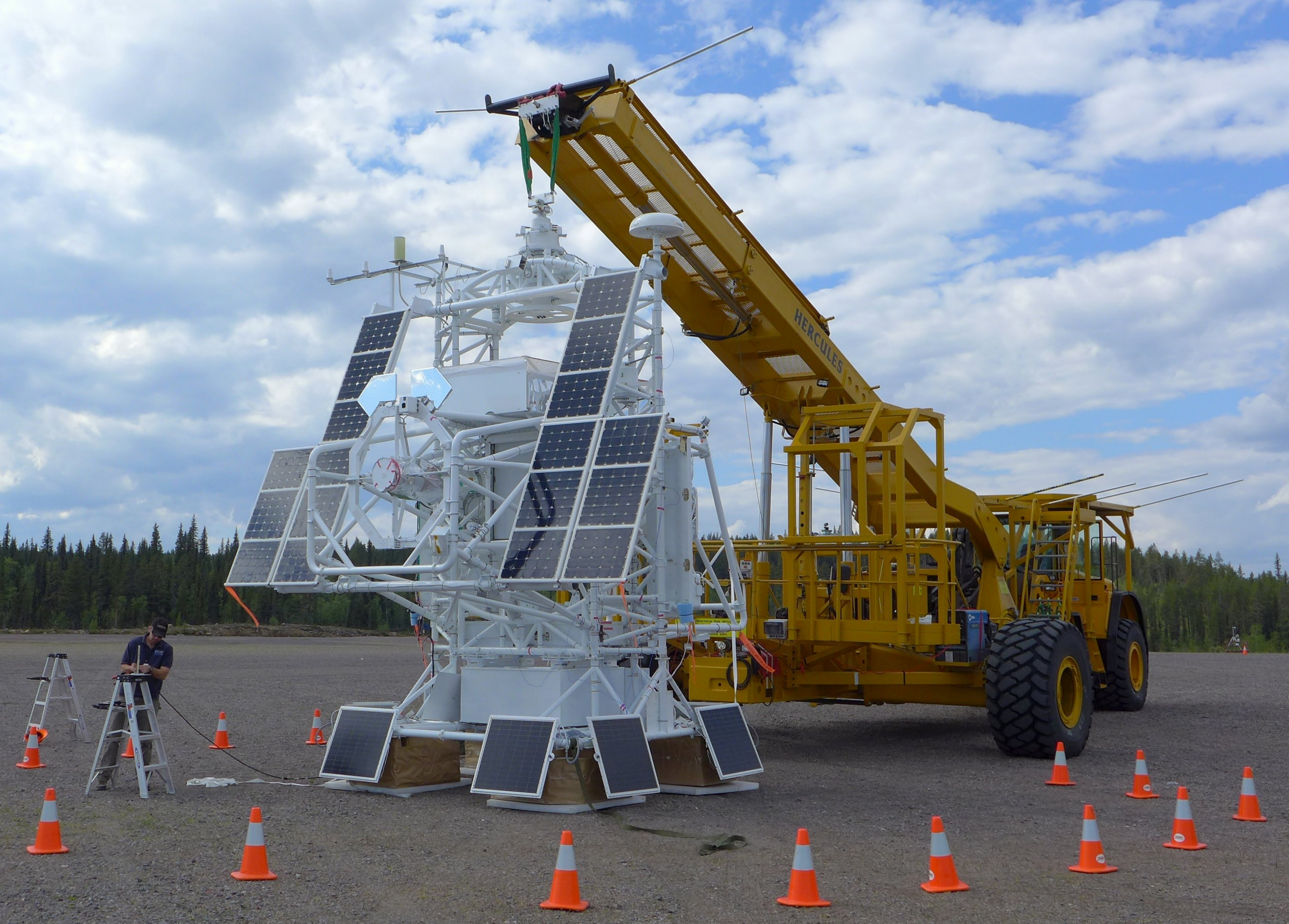}{0.45\textwidth}{(a)}}
\gridline{\fig{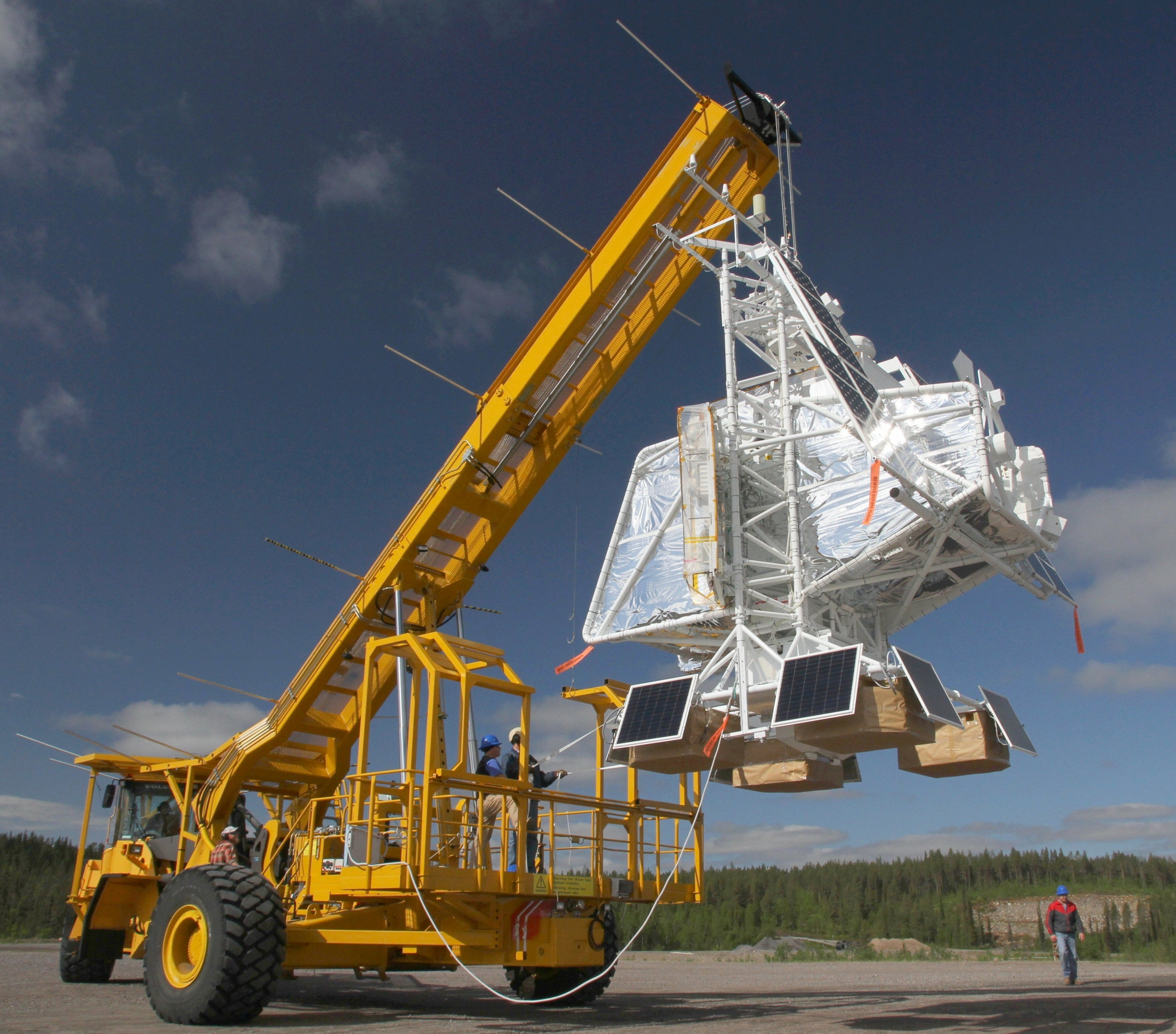}{0.45\textwidth}{(b)}}
\caption{Two views of the flight-ready \sunrise{}~II\ payload prior to the 2013 flight hanging
from the launch crane at ESRANGE. Image (a) shows the payload during
the electromagnetic compatibility test. The telescope front-ring can be seen
along with the radiators mounted above it. Multi-layer insulation (MLI) is used
to thermally insulate the
telescope and the instruments. Image (b) shows better the gondola structure, including
front and rear roll cages.}
\label{f1}
\end{figure}

Figure~\ref{f1} shows two views of the \sunrise{}~II payload (telescope with instruments
and electronics mounted in the gondola) hanging from the launch crane at the
ESRANGE balloon launch facility. More details about the whole system are
provided by \cite{Barthol2011}.

\subsection{Post-focus instrumentation}\label{subsec:instrum}

The instrumentation mounted within the Post-Focus Instrumentation platform (PFI)
survived the landing and recovery in good shape. Some cleaning was necessary,
but almost all elements could be re-used. The PFI was refurbished, but remained
technically identical to the one used during the 2009 flight. The scientific
instruments were also refurbished and subsequently integrated and realigned.

{\it ISLiD:}
The Image Stabilisation and Light Distribution system \citep{Gandorfer2011}
was not modified relative to the first flight. In order to optimise the
end-to-end optical performance of the combined ISLiD-IMaX path, an additional
plane-parallel plate was introduced into the converging IMaX feed path in front
of the IMaX interface focus. By choosing the inclination and orientation of this
plate, residual
astigmatism in the ISLiD/IMaX path could be minimised. The plate is coated
with high-efficiency anti-reflective layers on both sides, which are tuned
to be unpolarizing at the selected angle of incidence.

\begin{figure}
\figurenum{2}
\gridline{\fig{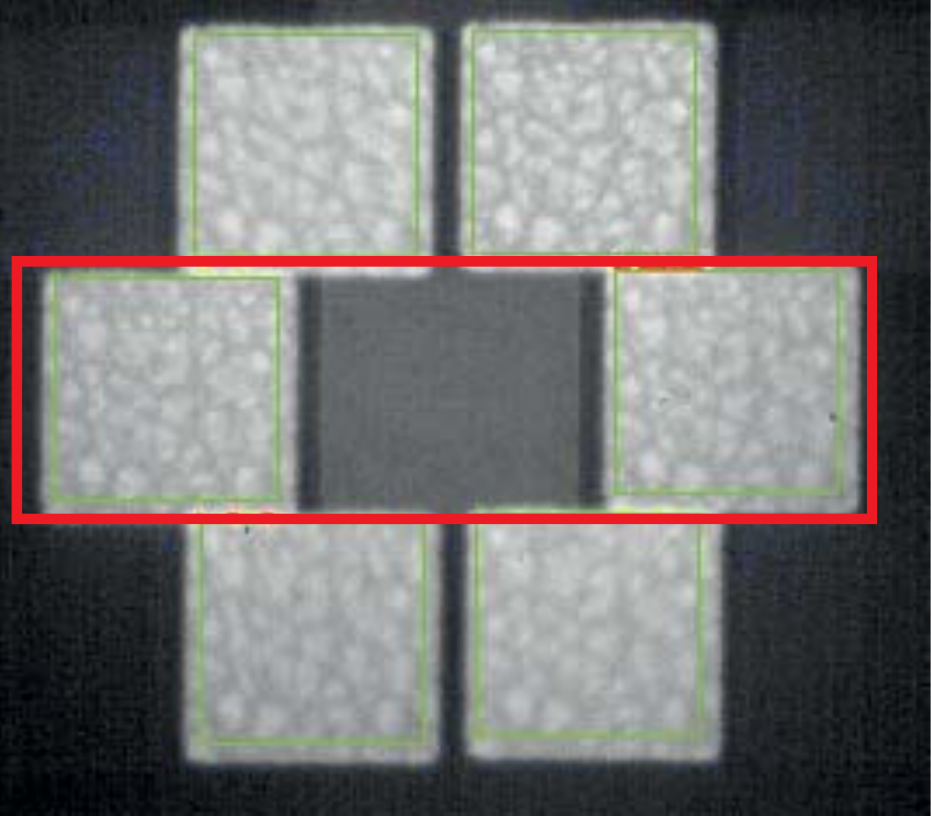}{0.4\textwidth}{}}
\caption{Snapshot of the image recorded by the Correlating
Wavefront Sensor (CWS), showing the 6 sub-apertures displaying the same solar
scene. The red frame bounds the two images employed in the 2-sub-aperture mode.}
\label{f2}
\end{figure}

{\it CWS:}
The main improvement to the Correlating Wavefront Sensor \citep{Berkefeld2011}
for the 2013 flight consisted of a second operating mode. By reading
out only the two sub-apertures in the center row (marked red in Fig.~\ref{f2}), the
bandwidth could be substantially increased and the residual image jitter decreased.
Table~\ref{tab:tableCWS} compares the two modes.

Due to the better performance, the two-sub-aperture mode was used
exclusively during the 2013 flight. Further software improvements included
better data logging capabilities and much faster focusing after closing the
CWS control loop.

\floattable
\begin{deluxetable}{lrr}
\tablecaption{CWS parameters for the six-sub-aperture and two-sub-aperture modes \label{tab:tableCWS}}
\tablehead{
\colhead{Parameter} & \colhead{6-sub-aperture} & \colhead{2-sub-aperture}\\
\colhead{} & \colhead{ mode} & \colhead{ mode}
}
\startdata
Control loop frequency [Hz]  &     1400          &       4000   \\
Bandwidth (0db) [Hz]         &      90           &        140   \\
Measurement accuracy [mas]   &       2           &          4   \\
Detectable Zernike modes     & tilt, focus, coma & tilt, focus  \\
Residual image jit-          &  50 (in 2009)     &   ---        \\
ter [mas RMS]                &  35 (in 2013)     & 25 (in 2013) \\
\enddata
\end{deluxetable}

{\it SuFI:}
The \sunrise{} Filter Imager \citep{Gandorfer2011} was changed very little
compared to the \sunrise{}~I flight. Thus, the mechanical shutter,
which had seen more than $150\,000$ releases during the 2009 flight, was
replaced. For the SuFI CCD
camera a new power supply unit with very low output noise was developed.
This led to an improvement in signal-to-noise performance by a factor of more
than ten.  In addition, three of the
wavelength filters were exchanged. The filter set for observations in the
2140~\AA\ region was replaced by a combination of two different filters,
both of which have a FWHM of 210~\AA, but have different side band characteristics,
which ensures sufficient blocking of longer wavelength radiation.
The 3120~\AA\ channel of \sunrise{}~I was replaced by a combination of a
2795~\AA\ filter with
4.8~\AA\ width and two blockers of 300~\AA\ and 110~\AA\ width, respectively.
This combination of filters, whose profile is plotted in Fig.~1 of \cite{Riethmueller2013b},
is centred on the Mg~{\sc ii}~k line and gets minimal contribution
from the Mg~{\sc ii}~h line located in the filter profile's wings.
Observations in the Ca~{\sc ii}~H line were possible through two different
filters centred at 3968~\AA. In addition to the 1.8~\AA\ wide filter already
used in \sunrise{}~I, a
1.1~\AA\ wide filter was available, which provided better isolation of the
contributions from higher atmospheric layers. In the following, these
channels are referred to as the 3968w ("w" for wide wavelength band) and
3968n ("n" for narrow band) channels. The 3880~\AA\ channel, which was used
in \sunrise{}~I, was sacrificed for this purpose.

{\it IMaX:}
The  Imaging Magnetograph eXperiment \citep{MartinezPillet2011a}
on the 2013 flight was also very similar to the version flown on \sunrise{}~I, although a
number of smaller changes and updates had been made. The parts
replaced included the FPGA in the proximity electronics, the LiNbO$_3$
etalon, one of the CCD cameras, the collimator doublet, the liquid crystal
variable retarders (LCVRs), the power supply for the
LCVR heaters (the heaters on the \sunrise{}~I flight did not have enough
power to bring the instrument to its nominal operational temperature) and
the phase-diversity plate (the original broke upon landing after the
\sunrise{}~I flight). Most of these parts were replaced by nearly
identical ones. In addition, the windows of both CCD cameras
were removed in order to avoid fringes that had been observed during the
\sunrise{}~I flight.

\section{Description of the mission} \label{sec:mission}

\begin{figure}
\figurenum{3}
\gridline{\fig{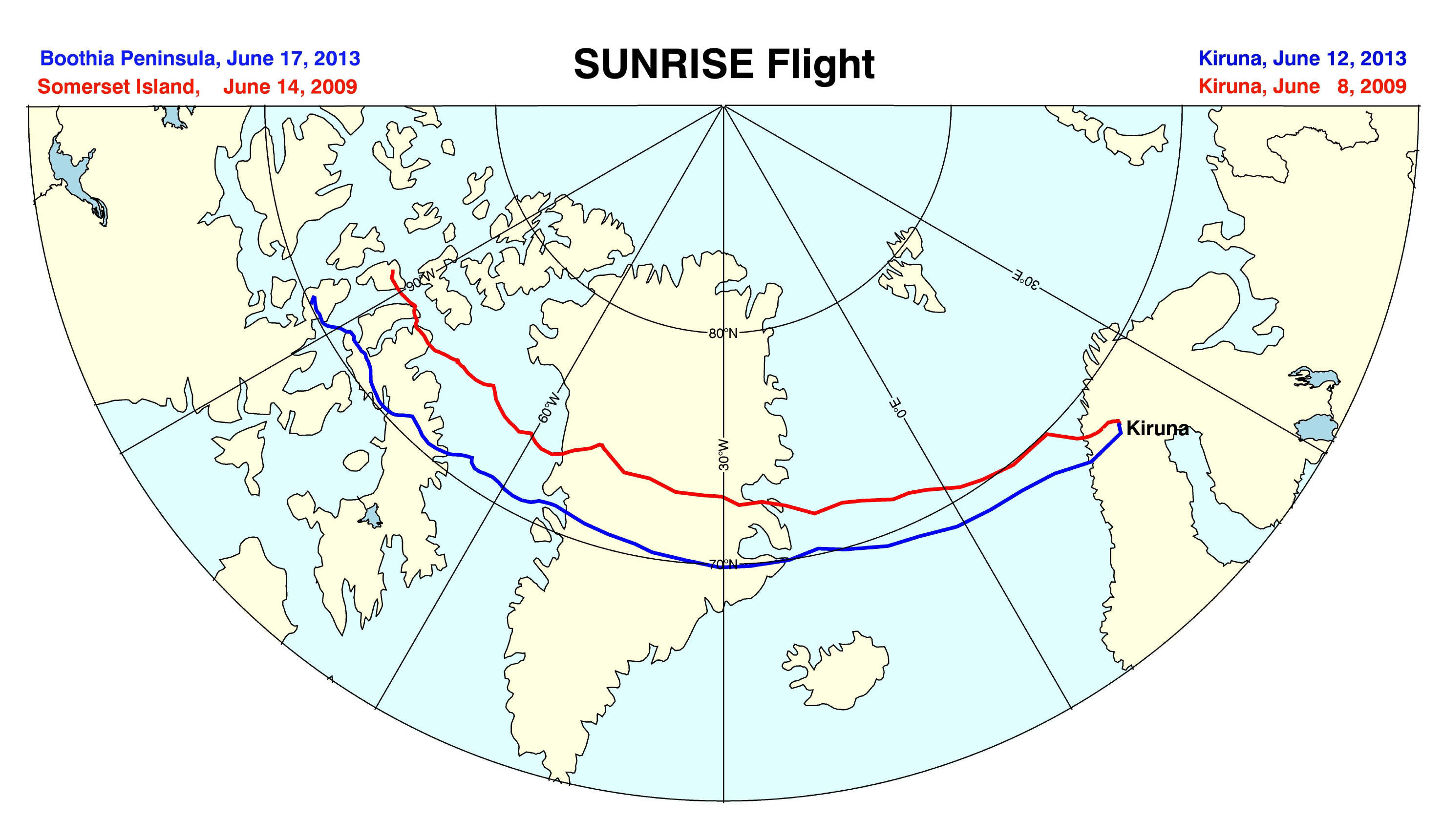}{0.48\textwidth}{(a)}}
\gridline{\fig{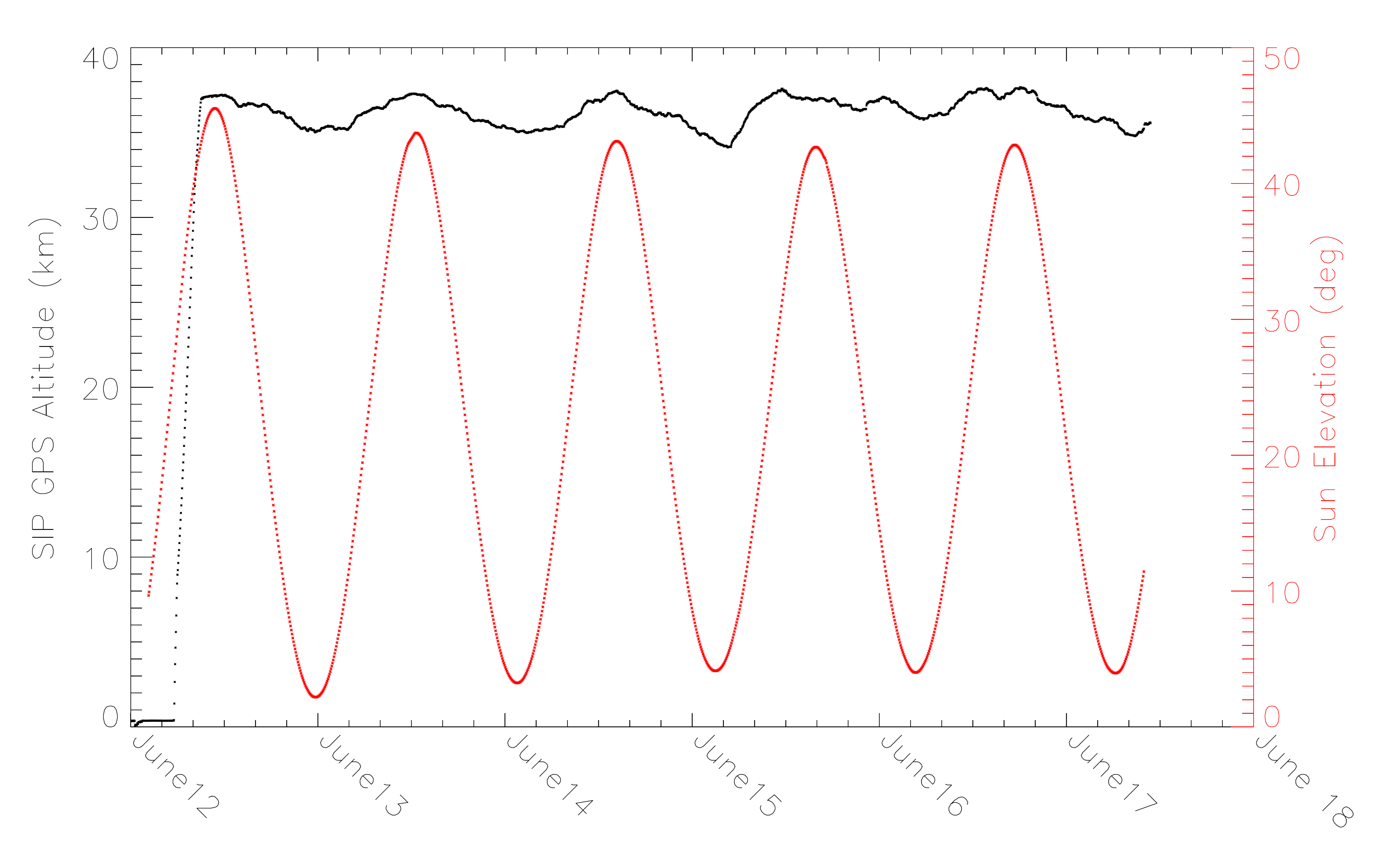}{0.48\textwidth}{(b)}}
\caption{(a) The flight paths of the 2009 (red curve) and 2013 (blue curve) \sunrise{}
science flights overplotted on a map of the northern Atlantic. The semi circles
mark latitudes of 60, 70 and $80^\circ$, respectively. (b) The \sunrise{}~II
float altitude vs. time as recorded by the CSBF-provided Support Instrumentation Package (SIP),
from shortly before launch up to the time of cut-off from the balloon (black curve, referring
to the left axis). Also plotted is the solar elevation angle, as seen from \sunrise{}~II
(in red, referring to the axis on the right).}
\label{f3}
\end{figure}

\sunrise{}~II was flown on a zero-pressure stratospheric long-duration balloon,
launched and operated by the Columbia Scientific Ballooning Facility (CSBF).
It was launched on June 12, 2013 at 05:37:53 UT (07:37:53 local time) from
ESRANGE (67.89$^\circ$N, 21.10$^\circ$E) near Kiruna in northern Sweden on a cloudy, but perfectly
windstill day. It reached a float altitude of 37.1~km after an ascent lasting
approximately 3.5~h. It then drifted westward at a mean speed of 35.3~km\,h$^{-1}$
and at a mean altitude of roughly 36~km.
Compared to \sunrise{}~I it travelled more to the south and somewhat faster
across the Atlantic and Greenland, so that it reached the peninsula of Boothia
in Northern Canada on June 17, 2013 where the flight was terminated
at 11:49:24 UT. After 122.3 hours at float altitude the balloon was cut off
and the payload descended suspended from a parachute, reaching the ground at
70.08$^\circ$N, 94.42$^\circ$W about one hour later. It landed
relatively softly but tipped over forward, so that the front ring of the
telescope and the radiators of the heat rejection wedge thermal control system
were damaged. Otherwise very little serious damage was suffered.
Figure~\ref{f3}a shows the flight path of \sunrise{}~II overplotted on a map. Also
plotted, for comparison purposes, is the flight path taken by \sunrise{}~I almost
exactly 4 years earlier. Figure~\ref{f3}b displays the height profile of the flight,
with the day-night cycles being visible (although the Sun never quite set on
the payload at float altitude). Rapid, although usually not very large increases in height, for
instance on June 15 starting at 05:04~UT, are due to ballast drops. The elevation of
the Sun as visible from \sunrise{}~II is overplotted. Clearly, the payload
remained in direct sunlight during the entire flight, although the Sun was just
over the horizon around local midnight.

Since at float altitude the payload was
above 99\%\ of the Earth's atmosphere, virtually seeing-free observations were
possible all the time. Also, the balloon stayed above most of the ozone in the
Earth's atmosphere, allowing high-resolution imaging in the UV at 2140~\AA,
2795~\AA\ and 3000~\AA, although the residual atmosphere did require excessively
long observing times at 2795~\AA\ and, to a smaller extent, at 2140~\AA.

E-Link worked until June 13, 2013 at 1:05 UT, i.e. until well after commissioning
was completed. Two events occurred after commissioning that impacted the
science operations.
1. The temperature controller of the IMaX etalon failed on June 12, 2013 at 23:55
UT, although possibly problems in the communications between IMaX subsystems
was the real reason for the failure.

2. On June 13, 2013 at 7:30 UT, about 23 hours after the observations had started,
the highly reflective front face of the heat rejection wedge --- a glued, 0.1\,mm
thin second surface mirror ---
failed. Due to increased absorption the heat rejection wedge temperature rapidly
increased each time the telescope was pointed at the Sun, eventually exceeding the
temperature sensor measurement range (whose upper limit lay at $130^\circ$ C)
within a number of minutes (the exact number depended somewhat on the time of day).
Therefore, all observing sequences had to be shorter than this interval after
this incident, with roughly half an hour in between observing sequences to allow
the heat rejection wedge to cool down again. However, the optical quality of
the data obtained after the failure of the heat rejection wedge remained the
same as before it, as a careful inspection revealed.

%

\section{Overview of the data recorded during the 2013 flight of \sunrise{}} \label{sec:overview}

The total length of time over which observations were made was 122 hours, during which
period SuFI recorded 300~GB (60806 images), while IMaX acquired 68~GB of data
(48129 images).
During 16\% of the total time at float altitude the CWS loop was closed.
The longest time series of SuFI data covers 60~minutes, while the longest
currently reduced IMaX time series lasts approximately 17~min.

SuFI observed in a variety of modes, differing mainly in the wavelengths
sampled. The various modes were recorded at different times of day.
Thus the two shorter wavelengths, at 2140~\AA\ and at 2795~\AA, were
only recorded close to local noon, when the Sun was the highest in the Sky
and the absorption due to the atmosphere was minimal.
A brief summary of the recorded SuFI data is given in Tables~\ref{SufiModes}
and~\ref{LongestTimeSeries}. The plate scale per pixel of SuFI depended
on the wavelength and was in the range 0.01983\arcsec-0.02069\arcsec.

IMaX data were dominantly obtained in the \mbox{V8-4} mode, meaning that
the full Stokes vector of the Fe~{\sc i} 5250.2 \AA\ line was
recorded at 8 wavelength positions, with 4
accumulations at each wavelength. The wavelengths sampled
were centred on $-120, -80, -40, 0, +40, +80, +120$, and $+227$~m\AA\ from
line centre. The last of these samples a continuum position between the
$g=3$ line and its
neighbouring Fe~{\sc i} line at 5250.6~\AA. Each exposure lasted 250~ms, so
that the cadence achieved with the IMaX \mbox{V8-4} mode was 36.5~s.
The plate scale per pixel of IMaX remained unchanged at 0.05446\arcsec.

A relatively small amount of IMaX data were also recorded in the
\mbox{L12-6} mode, in which only Stokes $I$ and $V$ are recorded at 12 wavelength
points in the line, with 6 accumluations per wavelength point.

\floattable
   \begin{table}
   \caption{SuFI observing modes}
   \label{SufiModes}                                     
   \centering                                            
   \begin{tabular}{l r}                                  
   \hline                                                
   \noalign{\smallskip}
   SuFI mode                   & Fraction of \\
                               & observing time \\
   \hline
   \noalign{\smallskip}
   {$5\lambda$: 2140, 2795, 3000, 3968w, 3968n~\AA}  & {54\%}    \\
   \noalign{\smallskip}
   {$4\lambda$: 2140, 3000, 3968w, 3968n~\AA}  & {2\%}    \\
   \noalign{\smallskip}
   {$3\lambda$: 3000, 3968w, 3968n~\AA}  & {33\%}    \\
   \noalign{\smallskip}
   {$2\lambda$: 3000, 3968w~\AA}     & {~8\%}    \\
   \noalign{\smallskip}
   {$1\lambda$: 3968w~\AA}     & {~3\%}    \\
   \noalign{\smallskip}
   \hline                                                
   \end{tabular}
   \tablecomments{3968w refers to the SuFI channel observed through the
   1.8~\AA\ wide filter centred on the line core of Ca~{\sc ii}~H, while
   3968n refers to the channel observed through the 1.1~\AA\ wide Ca filter}
\end{table}

\floattable
   \begin{table*}
   \caption{List of the longest SuFI time series of active region observations}
   \label{LongestTimeSeries}                                 
   \centering                                            
   \begin{tabular}{l l l l l l l l }
   \hline                                                
   \noalign{\smallskip}
   Start time      &  Duration & Filter                      & Exp. times          & Cadence  & $\mu$ & AR    & IMaX  \\
   \noalign{\smallskip}
   [UT]            &  [s]  & [\AA]                        & [ms]                & [s]      &       &       & mode  \\
   \hline
   \noalign{\smallskip}
   12.06. 23:39    & 3528  & 3000, 3968w, 3968n            &       500, 100, 500 &  7.2     & 0.93  & 11768 & V8-4  \\
   13.06. 01:10    & 741   & 3000, 3968w, 3968n            &       500, 100, 500 &  7.2     & 0.93  & 11768 & n.a.  \\
   17.06. 01:43    & 528   & 2140, 3000, 3968w, 3968n      & 2500,  50, 100, 500 & 10.8     & 0.45  & 11775 & V8-4  \\
   16.06. 04:37    & 470   & 3000, 3968w, 3968n            &       200, 100, 500 &  5.1     & 0.30  & 11775 & V8-4  \\
   12.06. 19:59    & 462   & 3000, 3968w, 3968n            &        50, 100, 500 &  6.0     & 0.94  & 11768 & L12-6 \\
   16.06. 00:43    & 353   & 2140, 3000, 3968w, 3968n      & 2000,  50, 100, 500 & 10.4     & 0.71  & 11770 & V8-4  \\
   17.06. 06:02    & 344   & 3000, 3968w, 3968n            &        50, 100, 500 &  5.1     & 0.43  & 11775 & n.a.  \\
   16.06. 05:04    & 335   & 3000, 3968w, 3968n            &       200, 100, 500 &  5.1     & 0.30  & 11775 & n.a.  \\
   16.06. 00:10    & 294   & 3000, 3968w                   &        50, 100      &  2.5     & 0.71  & 11770 & n.a.  \\
   16.06. 05:54    & 263   & 3000, 3968w, 3968n            &       200, 100, 500 &  5.1     & 0.29  & 11775 & n.a.  \\
   14.06. 02:58    & 259   & 1$\times$3000, 5$\times$3968w &       500, 100      &  7.6/1.2 & 0.37  & 11770 & n.a.  \\
   17.06. 02:37    & 214   & 3000, 3968w, 3968n            &        50, 100, 500 &  5.1     & 0.42  & 11775 & V8-4  \\
   \noalign{\smallskip}
   \hline                                                
   \end{tabular}
   \end{table*}


\section{Data reduction} \label{sec:reduction}
Due to the changes in the instrumentation as well as the experience gained
from \sunrise{}~I, the data were reduced using a modified reduction pipeline.

\subsection{SuFI}
The SuFI data, as on the first flight, were acquired in phase-diversity (PD) mode,
i.e., one half of the sensor was located in the focal plane,
while the other half imaged the same part of the solar surface as the first
half, but with a fixed offset in the focus direction. This configuration is necessary,
since aberrations of the telescope, although small enough to allow for diffraction
limited performance at visible wavelengths, are not negligible at the
much shorter UV wavelengths. The PD technique in principle allows for the
determination and subsequent removal of these aberrations, provided one of the
pair of images is in or very near focus according to \cite{Gonsalves1979};
\cite{Paxman1992}.

Therefore, after the traditional darkfield and flatfield corrections, the data
were restored using the Multi-Frame Blind Deconvolution (MFBD) wavefront
sensing code \citep{vanNoort2005}, which allows for an arbitrary PD term to be
present in the data. Due to the highly non-telecentric configuration of
the beam, however, the image scale of the defocused image was slightly
different from the image scale in the focal plane, so that the influence of the
tip-tilt components of the wavefront aberrations in the defocused channel
differed from that in the focal plane. This effect was accommodated by
restoring the frames in the ``calibration mode'' of the MFBD code, where all
fitted wavefront modes are constrained to be the same for the two channels,
with the exception of the tip-tilt and focus terms, that are allowed to vary,
but with the difference between the focus and defocus channels constrained
to have the same value for all PD pairs in the dataset. Due to the relatively
low frame rate of the SuFI camera, only four PD pairs, recorded in
approximately twenty seconds, could be combined to restore each frame, beyond
which the solar evolution started to degrade the result.

Although this method worked well on some of the data, it did not always restore
the data to the same quality. This is believed to be the result of residual
pointing errors during the relatively long exposure time, that blurred the image
in a way that is not consistent with the model, which assumes the point spread
function (PSF) to be produced by a single, static wavefront.
The azimuthally averaged Fourier power spectrum of the best frames,
however, shows power significantly above the noise floor, at wave numbers up to 85\%\
of the true diffraction limit of $D_{telescope}/lambda$ (DL, see for instance Paxman et al. (1996)),
which is slightly better than the criterion of $D_{telescope}/(1.22 \lambda)$ proposed by Rayleigh (1879)
and generally adopted for considering individual features to be resolved.

\subsection{IMAX}
 The reduction applied to the data from the 2009 flight was explained
in detail by \citet{MartinezPillet2011a}. Consequently, here we focus on
changes in the data reduction with respect to the reduction of the first flight's
data, necessitated by the replacement of several components of IMaX (described
in Sect.~\ref{subsec:instrum}). These replacements
influenced various instrumental effects of IMaX such as the occurrence of
fringes, ghost images, and the instrument's stray-light
behavior.

After the data were corrected for dark current and flat-field effects, residual
excess power due to interference fringes was filtered out in the Fourier domain.
Before and after each observing run during the flight, a PD measurement was
obtained by inserting a glass plate into the optical path of the first camera
in order to record pairs of defocussed and focussed images, which then allowed
retrieving the system's PSF. The observational data were reconstructed
by applying a modified Wiener filter constructed from the PD PSF. By
spatially replicating the images before performing operations in the Fourier
domain, a reduction of the FOV as a side effect of a necessary apodisation
could be avoided. Hence the effective FOV of the IMaX data from \sunrise{}~II
is $51\arcsec\times51\arcsec$.

The instrumental polarization was removed from the observations by a
demodulation matrix determined from the pre-flight polarimetric calibration. In contrast
to the first flight, this time the field-of-view dependence of that matrix was
taken into account. Since the on-ground polarimeric calibration did not include the
main mirror of the telescope and because the thermal environment during calibration
was different to the in-flight situation, cross-talk with Stokes $I$ had to be removed
from the data. Then the images of the two cameras were co-aligned and merged.

In addition, a slight periodic motion of the image was removed, that was not
present in the data of the first flight and correlated well with the
switching period of the heating power of the LCVRs, which were mounted near
one of the folding mirrors. We also added a further step to the data reduction pipeline
that interpolated the spectral scans with respect to time in order to compensate
for the solar evolution during the IMaX cycle time of 36.5\,s.

Finally, we determined the spatial mean Stokes~$I$ profile and subtracted 25\%
of the mean profile from the individual profiles to correct
the data for 25\% global stray light. This is equivalent to deconvolving the
IMaX data with a constant stray light not varying over the FOV.
Details of the stray-light properties of IMaX can be found in
\citet{Riethmueller2016}.

The physical quantities of the solar atmosphere were then retrieved from the
reconstructed Stokes images via the
Stokes-Profiles-INversion-O-Routines \citep[SPINOR;][]{Frutiger2000a,Frutiger2000b},
which uses the STOPRO routines to solve the Unno-Rachkovsky equations
\citep{Solanki1987}.
A relatively simple inversion strategy was applied to get robust results: Three
optical depth nodes for the temperature (at $\log\tau=-2.5, -0.9, 0$)
and a height-independent magnetic field vector, line-of-sight velocity and
micro-turbulence. The synthetic spectra were not only calculated for the
Fe\,{\sc i} 5250.2~\AA\ line but also for its neighboring lines (Co\,{\sc i} line
at 5250.0~\AA\ and the Fe\,{\sc i} line at 5250.6~\AA) to also include
their influence. The spectral resolution of IMaX was taken into account by
convolving the synthesized spectra with the spectral transmission profile of
IMaX measured in the laboratory during a pre-flight calibration campaign
\citep[see bottom panel of Fig.~1 in][]{Riethmueller2014}.

The SPINOR inversion code was run for ten iterations, then the output maps of
the individual parameters were spatially
smoothed and used as the initial guess parameters to a further run of ten
iterations. The intermediate smoothing of the free parameters of the inversion
was introduced to get rid
of spatial discontinuities in the physical quantities that can
occur if the inversion gets stuck in local minima of the merit function at
individual pixels or groups of pixels. This procedure
was repeated five times (implying 50 iterations in all), with a gradually
decreasing amount of smoothing.
Finally, the resulting LOS velocity maps were corrected for the etalon
blueshift, which is caused by the collimated setup
\citep[see][]{MartinezPillet2011a}.

\section{Examples of data and science results} \label{sec:results}

\subsection{NOAA AR~11768}

The longest time series obtained by \sunrise{}~II was
focussed on NOAA AR~11768 at $\cos\theta=\mu=0.93$, where $\theta$ is the
heliocentric angle. The
observations were made on 2013, June 12, starting at 23:39~UT, about 1.5 days
after the initial appearance of the active region, at a time when magnetic flux was
still emerging. At that time the active region had developed a full-fledged
leading polarity sunspot, while the following polarity was mainly concentrated
in a string of pores.

\begin{figure*}
\figurenum{4}
\gridline{\fig{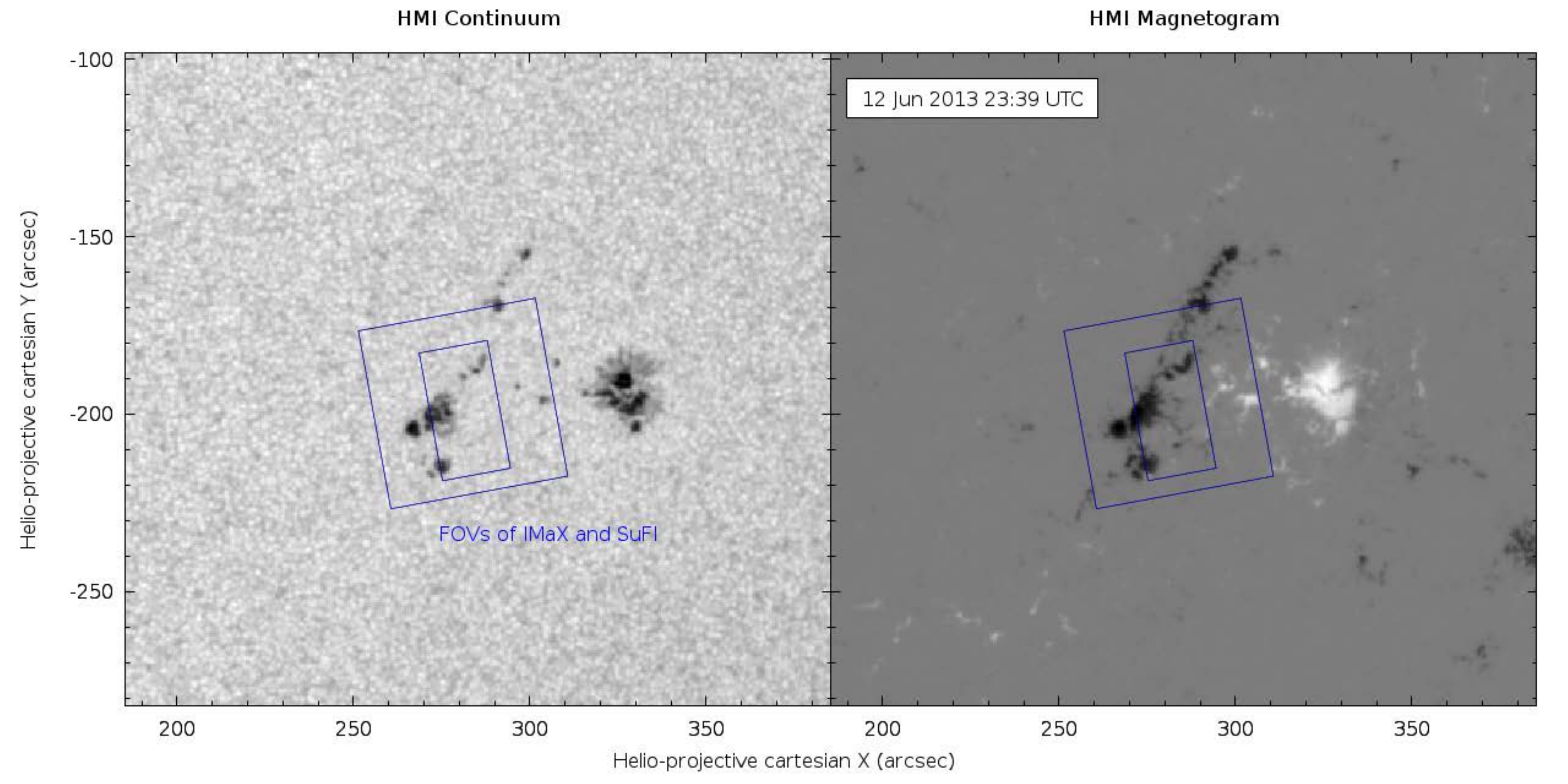}{1.0\textwidth}{}}
\caption{HMI continuum map (left image) and HMI magnetogram (right image)
recorded on 2013 June 12 at 23:39 UT, at the beginning of the \sunrise{}~II time
series. The x and y scales are in helio-projective
cartesian coordinates and are in arcsec, with
the origin located at solar disc centre. The outer blue box indicates roughly
the FOV of the IMaX instrument, while the inner blue box does the same for the
SuFI instrument. The animation of this figure shows the evolution of the active
region for 3 days around the \sunrise{}~II observations.}
\label{f4}
\end{figure*}

The active region is illustrated in Fig.~\ref{f4} at the time of the
\sunrise{}~II
observations. The IMaX and SuFI fields of view (FOVs) are overplotted on
an HMI continuum image and an HMI magnetogram and are indicated by the
blue boxes.\footnote{The SuFI FOV in Fig.~\ref{f4} is the maximum that can be
retrieved from the detector. The width of the images shown in Fig.~\ref{f5} is
lower for two reasons: if the broadest possible FOV is used, then the quality
of the reconstruction is lower and hence the resolution is less high than in
the images displayed in Fig.~\ref{f5}. Furthermore only the common FOV of the
various SuFI wavelength channels is plotted in Fig.~\ref{f5}. Due to
differential offsets, this is smaller than the FOV of each individual
channel.} IMaX covered most of the
following magnetic polarity pores, including the largest one. The largest pore
had a complex structure and on one side showed a feature with a roughly
penumbral brightness exhibiting elongated structures. In the course of the
further development, however, this feature did not develop into a proper
penumbra (as can be deduced from the animation of Fig.~\ref{f4}).
The IMaX FOV also contained some of the leading polarity flux, mainly in the
form of facular magnetic elements. \sunrise{}~II also caught a region of
emerging flux, to be found in the central southern part of the IMaX FOV and
partly in the much smaller SuFI FOV, which captured mainly the region between
the two polarities, including one of the emergence
events.

The animated Fig.~\ref{f4} shows the evolution of the active region from the
very first emergence 1.5 days prior to the start of \sunrise{}~II
observations of this region to approximately 1.5 days after the end of \sunrise{}~II
observations of this region. A significant amount of flux appeared
in the day after the \sunrise{}~II observations, which led to the formation of
a proper following polarity sunspot, as well as to an increased size of the
leading polarity spot, mainly through the coalescence of pores, many of which
already carried a piece of penumbra on one side prior to merging. The blue
boxes overlaid on the images in the animated figure are tilted by the rotation angle
relative to solar North and the roughly $5^\circ$ of rotation during the course
of the SuFI time series is taken into account. Note that the animated Fig.~\ref{f4}
runs at a slower speed during the time of the \sunrise{}~II observations of this region.

\subsection{Sample SuFI data}

\begin{figure*}
\figurenum{5}
\gridline{\fig{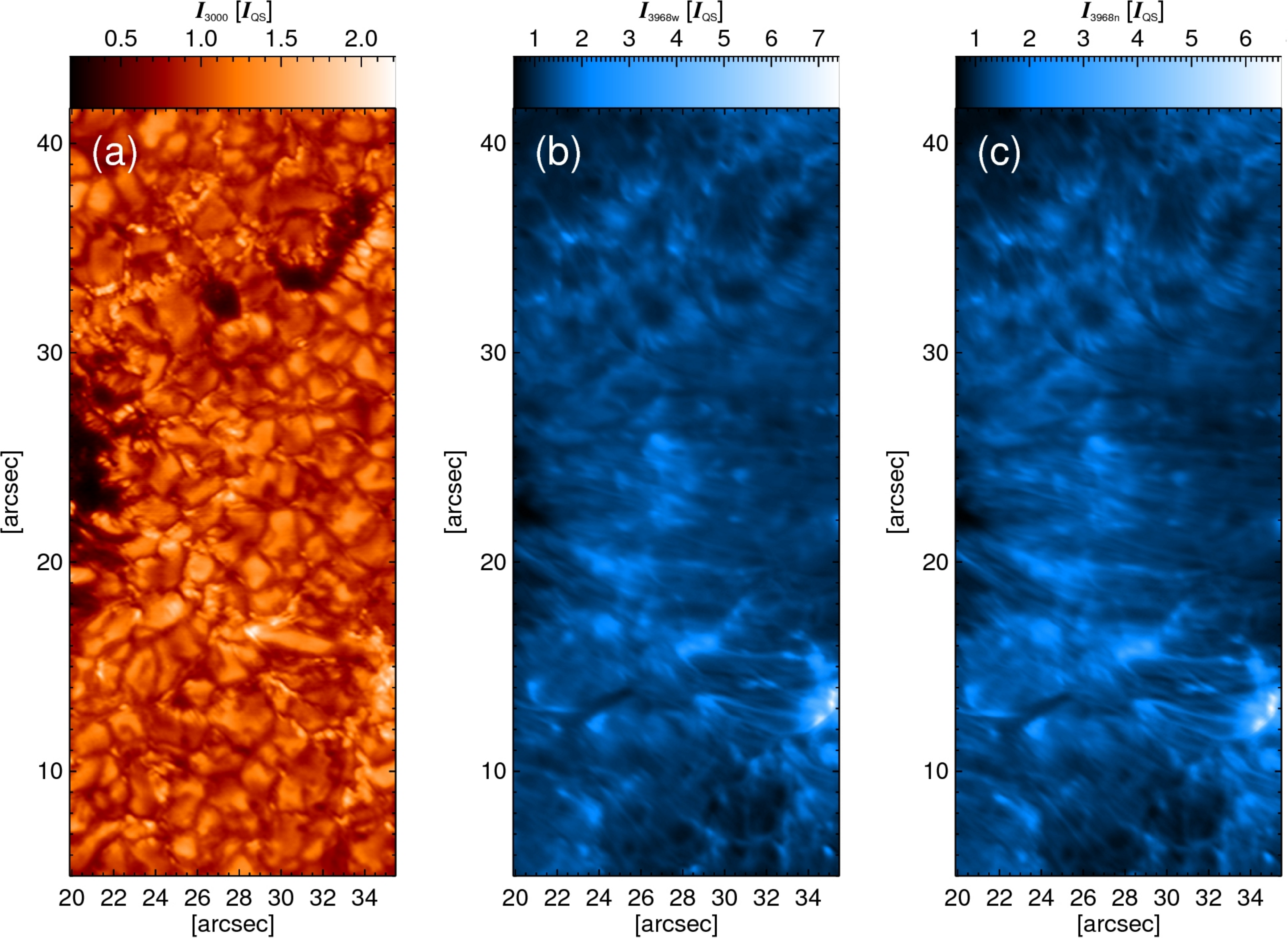}{0.8\textwidth}{}}
\caption{Images recorded by SuFI on 2013 June 12 at 23:46 UT after
multi-frame blind deconvolution incorporating phase diversity.
Plotted are (a) the intensity in the filter centred on 3000~\AA,
(b) intensity in the broad Ca~{\sc ii}~H channel and, (c) in the narrow
Ca~{\sc ii}~H channel. All intensities are given in units of the mean
quiet-Sun intensity. The coordinates are given with respect to the lower left
corner of the IMaX images (see Fig.~\ref{f6}). Note that in this figure
the plots at various wavelengths are aligned with each other
to an accuracy of only roughly 1\arcsec{}. }
\label{f5}
\end{figure*}

The SuFI instrument during the second flight of \sunrise{} provided
diffraction-limited images at 3000~\AA\ and at
3968~\AA\ in both the broader and the narrower channel centred on the core of
the Ca~{\sc ii}~H line.
An image at each of these wavelengths is plotted in Fig.~\ref{f5}.
Note that during this, the longest time series obtained by SuFI, no data in
Mg~{\sc ii}~k, nor in the 2140~\AA\ channel were obtained due to the low solar
elevation at that time and the consequent high atmospheric absorption at these
wavelengths.

Strongly different structures are seen at 3000~\AA\ and in the Ca~{\sc ii}~H
line, with the shorter wavelength
displaying granulation \citep[as had already been noticed in the \sunrise{}~I
data, e.g.,][]{Solanki2010, Hirzberger2010}, but also bright points
\citep{Riethmueller2010}, pores with internal fine structure and a bright
elongated granule located at approximately 30--32\arcsec\ in x-direction and
15--16\arcsec\ in the y-direction. This last feature is found to be associated
with magnetic flux emergence. The two brightest features in the 3000~\AA\ image
are located near the two ends of this structure, on its left it brightens to
the level of a strong bright point, while there is a bright facular region
right at the edge of the FOV in the continuation of the elongated granule on
its right side.

The surprising lack of difference between the broader and narrower Ca~{\sc ii}~H
channels in Fig.~\ref{f5} suggests that they were observing very similar
layers of the Sun. Possibly, the filter selecting the narrower channel had
drifted in wavelength (e.g. due to temperature changes in the instrument) and
was not centred exactly on Ca~{\sc ii}~H$_3$. Due to the similarity between the
broad and the narrow Ca~{\sc ii}~H channels, in the following the discussion
will concentrate on the narrower channel.

In the Ca~{\sc ii}~H images the dominant structures are slender fibrils directed roughly
across the SuFI FOV, many of which seem to emanate from the large pore just
to the left of the SuFI image. They appear to be similar to the
ones observed earlier by \cite{Pietarila2009}. Although most of the
filaments seem to roughly follow the same direction, a number of them do
cross each other. This is particularly well seen close to the location of the
emerging magnetic flux  mentioned above, at around (30--34\arcsec, 13--15\arcsec).
Very likely this is due to fibrils belonging to previously existing magnetic
field overlying the fibrils
associated with the emerging flux that are directed differently.
The lengths of these fibrils are hard to determine, due to inhomogeneities
that may be present in the background, but some seem to extend over a fair
portion of the width of the SuFI images. Strikingly, the fibrils are also
constantly evolving and in motion. The lengths,
along with other properties of these fibrils, are studied in a further
paper in this special issue \citep{Gafeira2016a}, while the dynamics of the
fibrils, in particular wave modes travelling along them, are investigated by
\citet{Jafarzadeh2016b} and by \citet{Gafeira2016b}.

The brightenings seen at 3000~\AA\ in
connection with the flux emergence are also visible in Ca~{\sc ii}~H. The
magnetic flux emergence caught by IMaX and SuFI is investigated
by \citet{Centeno2016} and \citet{Danilovic2016}. At a later
stage during the time series this brightening develops further into a small
flare that engulfs the emerging flux region and is prominently visible in the
Ca~{\sc ii}~H line.

\begin{figure}
\figurenum{6}
\gridline{\fig{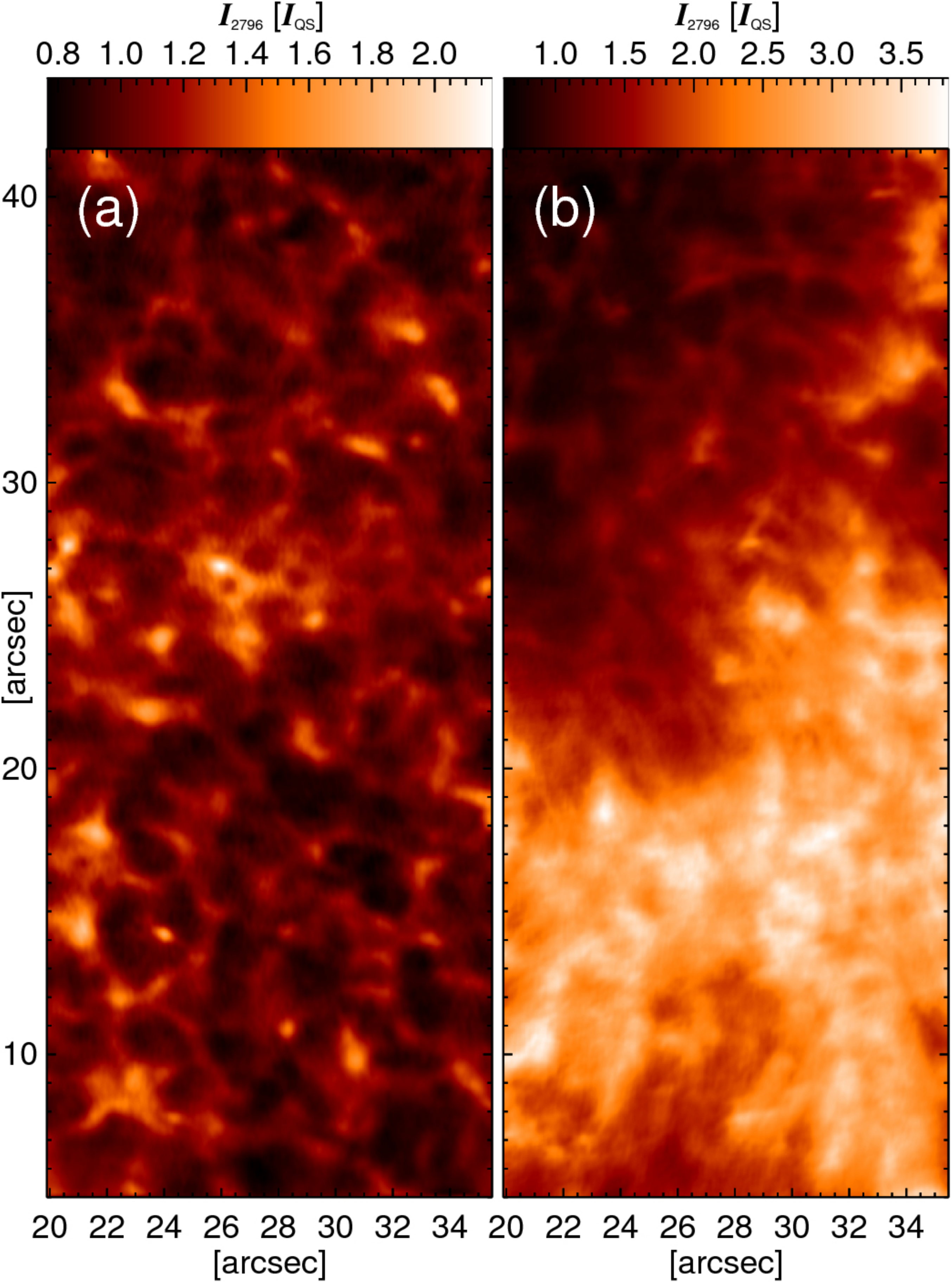}{0.4\textwidth}{}}
\caption{Images recorded by SuFI in the Mg II filter. Panel (a) displays a part
of the quiet Sun recorded on 2013 June 13 at 12:52:50 UT, while in Panel (b)
weak plage is shown that was observed on 2013 June 12 at 12:50:48 UT. Both
images were recorded exactly at disc centre and correspond to SuFI
level~3 data (see main text for details) and have been phase-diversity
reconstructed using an averaged point spread function.}
\label{fNEW}
\end{figure}

SuFI on board \sunrise{}~II also obtained the first high-resolution
images in the Mg~{\sc ii}~k line, which have already been published in two
early papers \citep{Riethmueller2013b,Danilovic2014}. Images taken in
the Mg~{\sc ii}~k filter were found to have considerable similarity with nearly
simultaneously recorded Ca~{\sc ii}~H images, although some distinct
differences were also found. These included a 1.4--1.7 times
larger intensity contrast and a more smeared appearance.
Exampls of Mg~{\sc ii}~k images of quiet Sun and of a weak plage/enhanced network
region are shown in Fig.~\ref{fNEW}. These images were exposed for 50 s to
counter the strong atmospheric absorption at this wavelength and to obtain a
sufficient S/N ratio. The images are based on SuFI level~3 data,
i.e. data that have been phase-diversity reconstructed using wave-front errors
averaged over the whole time series of images. See \citet{Hirzberger2011} for
more details on reduction and phase-diversity reconstruction of SuFI data.

It was concluded that some of these differences may be caused by the much
longer integration time needed to record the Mg~{\sc ii}~k images due to the strong
ozone absorption at that wavelength, while others are likely intrinsic.
Possible reasons for the intrinsic differences given by \citet{Danilovic2014}
include greater formation height and greater formation height range of the
Mg line (the latter due to the rather broad Mg filter) and the stronger response
of the emission peaks of Mg~{\sc ii}~k to temperature.

\subsection{Sample IMaX data and inversion results}

\begin{figure*}
\figurenum{7}
\gridline{\fig{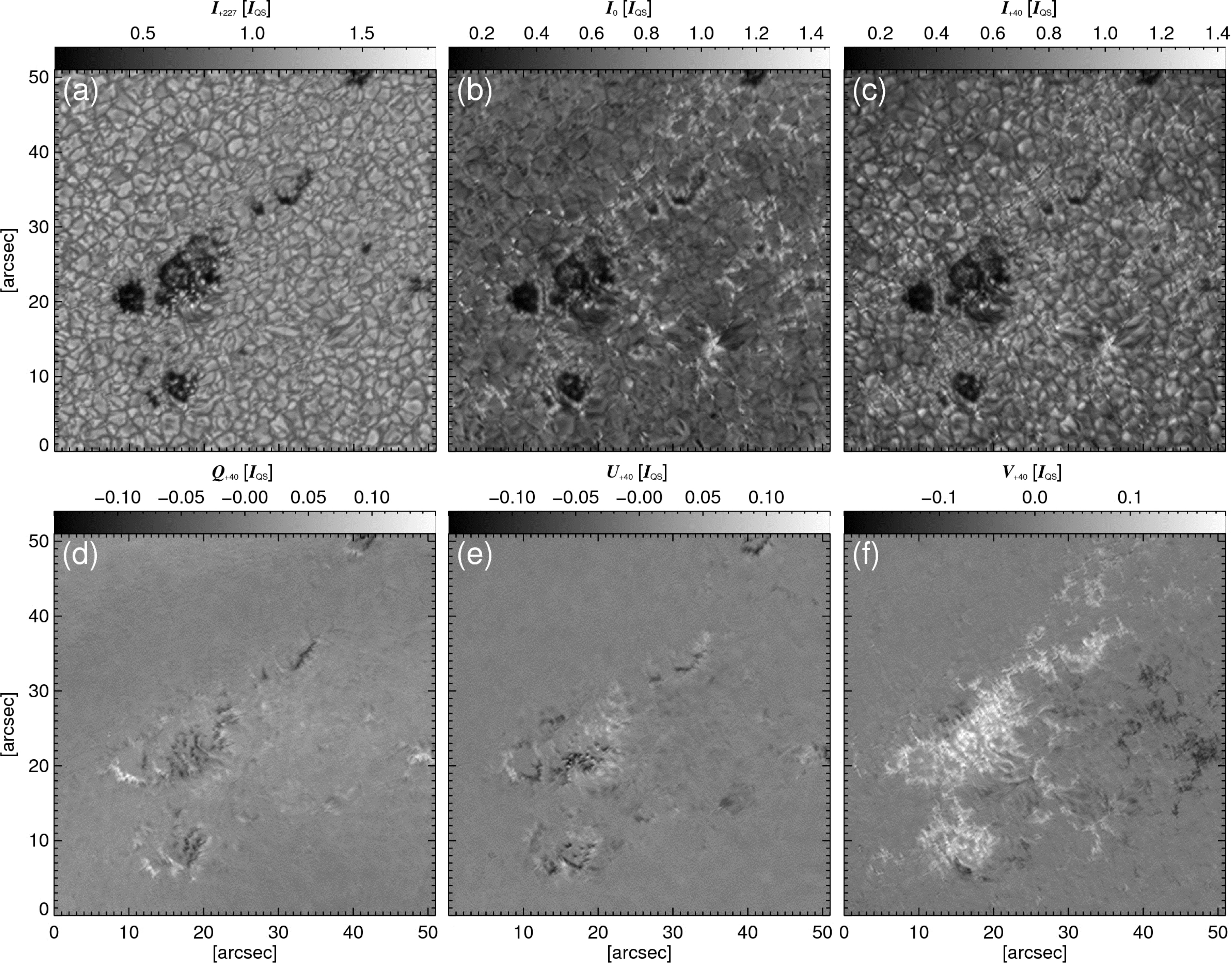}{1.0\textwidth}{}}
\caption{IMaX observables recorded on 2013 June 12 at
23:46 UT after phase-diversity reconstruction. Plotted are (a) the
continuum intensity, $I_{+227}$, (b) line core intensity (i.e. intensity in the wavelength
channel nominally located at the line core), (c)--(f) the
Stokes $I$, $Q$, $U$ and $V$ values at $+40$ m\AA\ from line centre. All
quantities are given in units of the respective quiet-Sun intensity, $I_{\rm QS}$.}
\label{f6}
\end{figure*}

The continuum intensity in the visible recorded by IMaX at $+227$~m\AA\ from the
line centre is plotted in Fig.~\ref{f6}a. IMaX images contain a larger number
of pores than SuFI due to the bigger FOV. As expected for
the continuum intensity at 5250~\AA, bright points are much less prominent
than at 3000~\AA\ (Fig.~\ref{f5}a),
although the fine structure in the largest pore is well visible, with some of the
bright ``umbral dots'' being among the brighter structures in the image.
Also well visible in that figure is the long elongated granule already
pointed out
in the 3000~\AA\ image (located at (30\arcsec,15\arcsec) in the IMaX FOV). Another
similar
granule is found to its right. Both are associated with magnetic flux emergence.

The bright points in the line core image displayed in Fig.~\ref{f6}b,
often forming connected chains filling the intergranular lanes, possess a high
contrast. Here, what we refer to as the line core intensity is simply
the intensity at the central IMaX wavelength point, which is nominally located
at the wavelength of the Fe~{\sc i} 5250.2~\AA\ line core.
Chains of such bright points surround most of the pores, but the
most prominent bright feature in this image is the strong brightening located
between the two elongated granules, i.e. between the two small flux emergence
events. The line core of Fe~{\sc i} 5250.2~\AA\ obviously forms below the
height of formation of the radiation sampled by either of the two SuFI
Ca~{\sc ii}~H filters, as the iron line core image still shows faint vestiges
of granulation and no signs of fibrils. All quantities in Fig.~\ref{f6} are
normalized to the respective intensity averaged over two areas within the joint
IMaX and SuFI FOV with near-quiet Sun conditions, i.e. a relatively low
magnetic flux density. We call this averaged ``quiet-Sun'' intensity,
$I_{\rm QS}$.  Note that the granule pattern changes
to that of reversed granulation if instead of being divided by $I_{\rm QS}$,
as is the case in Fig.~\ref{f6}b, the line core intensity is divided by the
local continuum intensity, i.e. $I_{+227}$, at each pixel \citep{Kahil2016}.

Stokes $I$, $Q$, $U$ and $V$ in the red flank of 5250.2 are
plotted in Figs.~\ref{f6}c--f. The differences between Fig.~\ref{f6}c
and Figs.~\ref{f6}a and b are due to the  formation height of the line flank
(at $+40$~m\AA) lying between
those of the other two wavelengths and the fact that the intensity in the line
flank, i.e., in Fig.~\ref{f6}c, is very sensitive to the Doppler shift of
the spectral line. The intensity in the line core, as plotted in
Fig.~\ref{f6}b, is also affected by Doppler shifts, but to a much lesser extent
than the line flanks.

\begin{figure*}
\figurenum{8}
\gridline{\fig{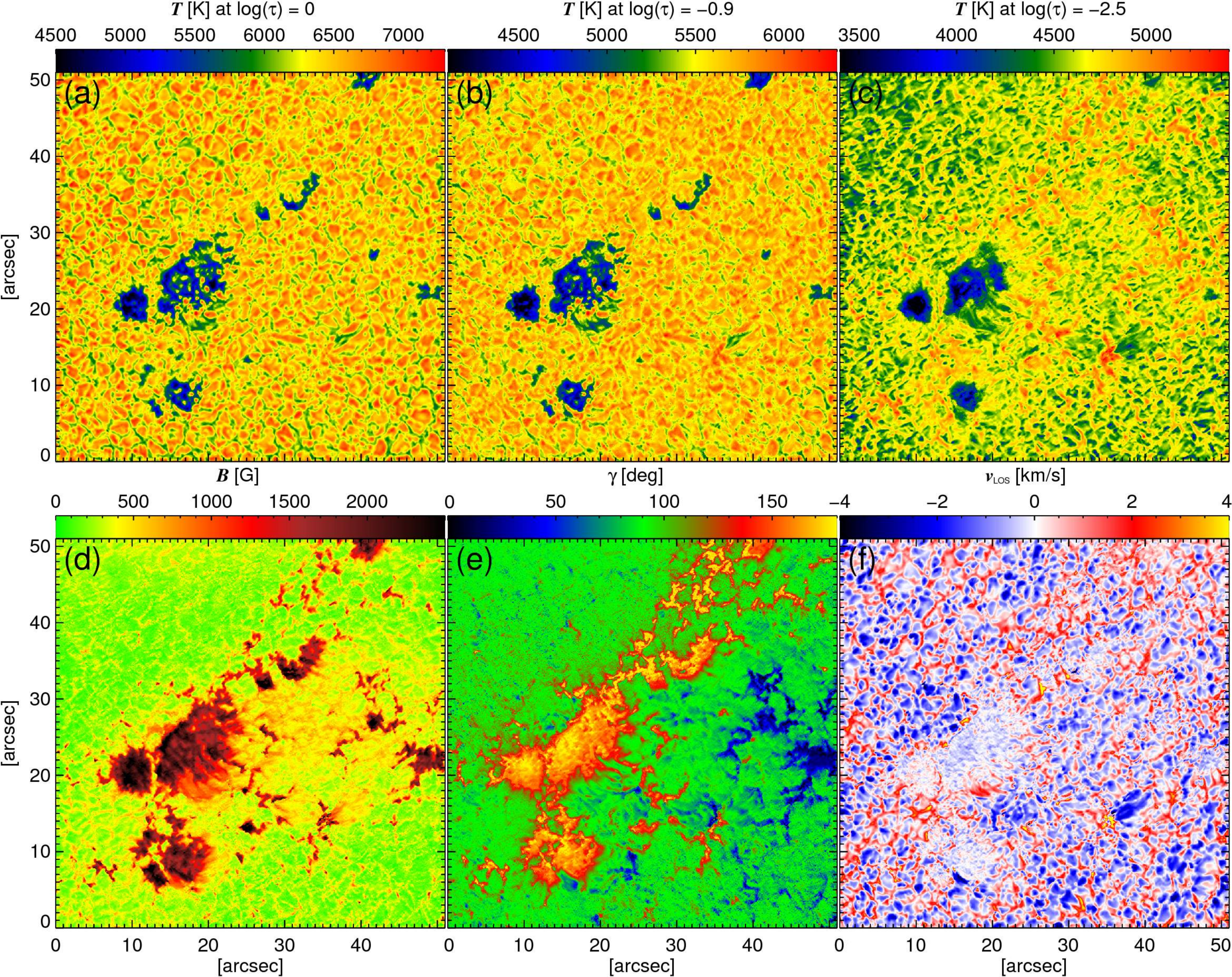}{1.0\textwidth}{}}
\caption{Best-fit atmospheric parameters deduced from the inversion of
the Stokes vectors recorded by \sunrise{}/IMaX and shown in Fig.~\ref{f6}.
Upper row of panels, (a)--(c): temperature $T(\log\tau=0)$,
$T(\log\tau=-0.9)$ and $T(\log\tau=-2.5)$.  Lower row, (d)--(f):
magnetic field strength, $B$, magnetic field inclination, $\gamma$,
and the line-of-sight velocity, $v_{\rm LOS}$. Postive velocities indicate
downflows. $\gamma$ is relative to the line-of-sight direction.}
\label{f7}
\end{figure*}

Maps of the best-fit parameters obtained from the inversion of the IMaX Stokes
vectors illustrated in Fig.~\ref{f6}, are presented in Fig.~\ref{f7}. The panels
in the upper row display the temperature at the three nodes at which it was
determined,
$\log(\tau)=0$, $\log(\tau)=-0.9$ and $\log(\tau)=-2.5$, the lower row of panels
gives the magnetic field strength, $B$, the inclination of the magnetic field
relative to the line-of-sight, $\gamma$, and the line-of-sight velocity, $v_{\rm
LOS}$.

The results of the inversions reveal some conspicuous features. Among these
are the strong
blue- and redshifts in the granulation (note that we have not removed the p-modes
from the inverted data, so that the velocities are influenced by such oscillations).
The velocity amplitude is larger than in the granulation shown by
\cite{Solanki2010}. An important reason for this is that scattered light has been
removed from the present data, while the data analysed by \cite{Solanki2010}
were still affected by scattered light. Some discrepancy can also be introduced
because the velocities displayed by \cite{Solanki2010} were obtained from a
Gaussian fit to line profiles sampled at 4 wavelengths, while the velocities
presented here were provided by the SPINOR inversions of line profiles
sampled at 7 wavelength points. One of the emerging flux patches is
found to be associated with a strong blueshift, much stronger than the other
emerging flux feature.

\subsection{Pores}

One noticeable feature of the inversion results is that the photospheric
field strength in pores often reaches 2500~G
\citep[compareble with the highest values found from Hinode data after
deconvolving with the PSF; ][]{QuinteroNoda2016}. In some places higher field strengths
are reached, but
these cannot be trusted due to the limited wavelength range of the observations
and the fact that the continuum point at $+227$ m\AA\ starts to be affected.

Another striking feature is that whereas all pores are clearly
cooler than their surroundings at the two lower heights ($\log\tau=0, -0.9$),
only the larger pores are clearly distinguishable from their surroundings in
the temperature map at
$\log\tau=-2.5$. The smaller pores can hardly be separated from their
surroundings. In contrast, the line core image in Fig.~\ref{f6} shows a
darkening also at the locations of the smallest pores. However, this is
mainly  due to the lower continuum intensity at these locations. Also, the line core is
affected by the Zeeman effect and to a smaller extent also Doppler shifts, so
that it is not an unalloyed measure of the
temperature. In order to test whether the inversions are giving a reasonable
representation of the temperature stratification in pores we consider the
SuFI Ca~{\sc ii}~H channel.

To compare SuFI and IMaX data the
images at individual SuFI wavelengths were aligned to sub-pixel accuracy
with the most similar IMaX images. For 3000~\AA\ this is the 5250.4~\AA\ continuum,
while the Ca~{\sc ii}~H channel is compared with the IMaX line core,
normalized to $I_{+227}$ at each pixel individually (then the images
look more similar). More information on
the procedure are provided by \cite{Kaithakkal2016}, cf.\ \cite{Jafarzadeh2016b}.

\begin{figure*}
\figurenum{9}
\gridline{\fig{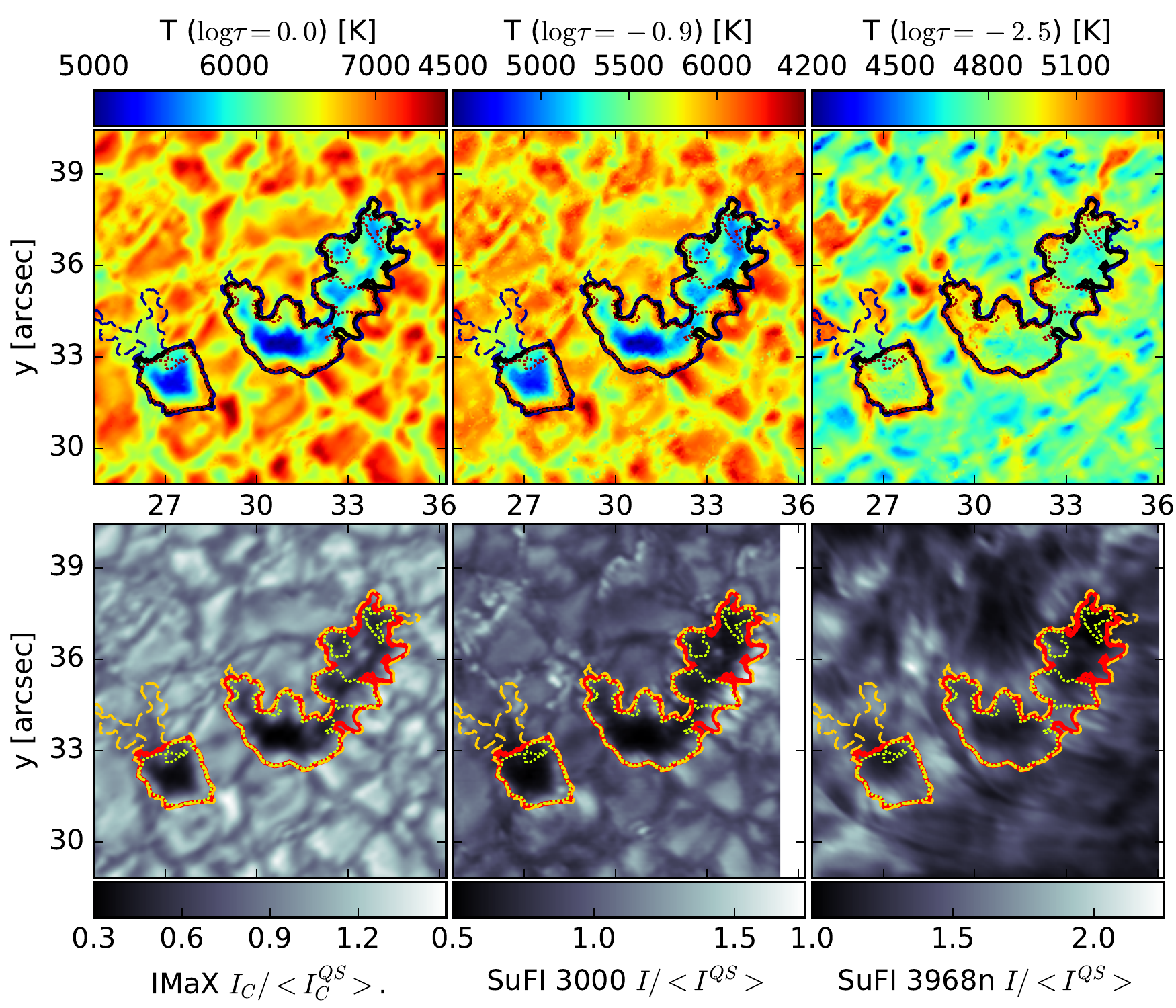}{0.8\textwidth}{}}
\caption{Blow-ups of a region containing two pores within the SuFI FOV.
Upper row of panels, from left to right: Temperature $T(\log\tau=0)$,
$T(\log\tau=-0.9)$ and $T(\log\tau=-2.5)$.  Lower row, from left to right:
IMaX continuum intensity, $I_{+227}$, SuFI 3000~\AA\ and SuFI 3968n channel.
The solid contours around the pores mark a field strength of 1400~G. For
illustration purposes, we have added contours corresponding to 1200~G (dashed)
and 1600~G (dotted). In the lower panels the contours have been given different
colours to enhance clarity.}
\label{f8}
\end{figure*}

\begin{figure*}
\figurenum{10}
\gridline{\fig{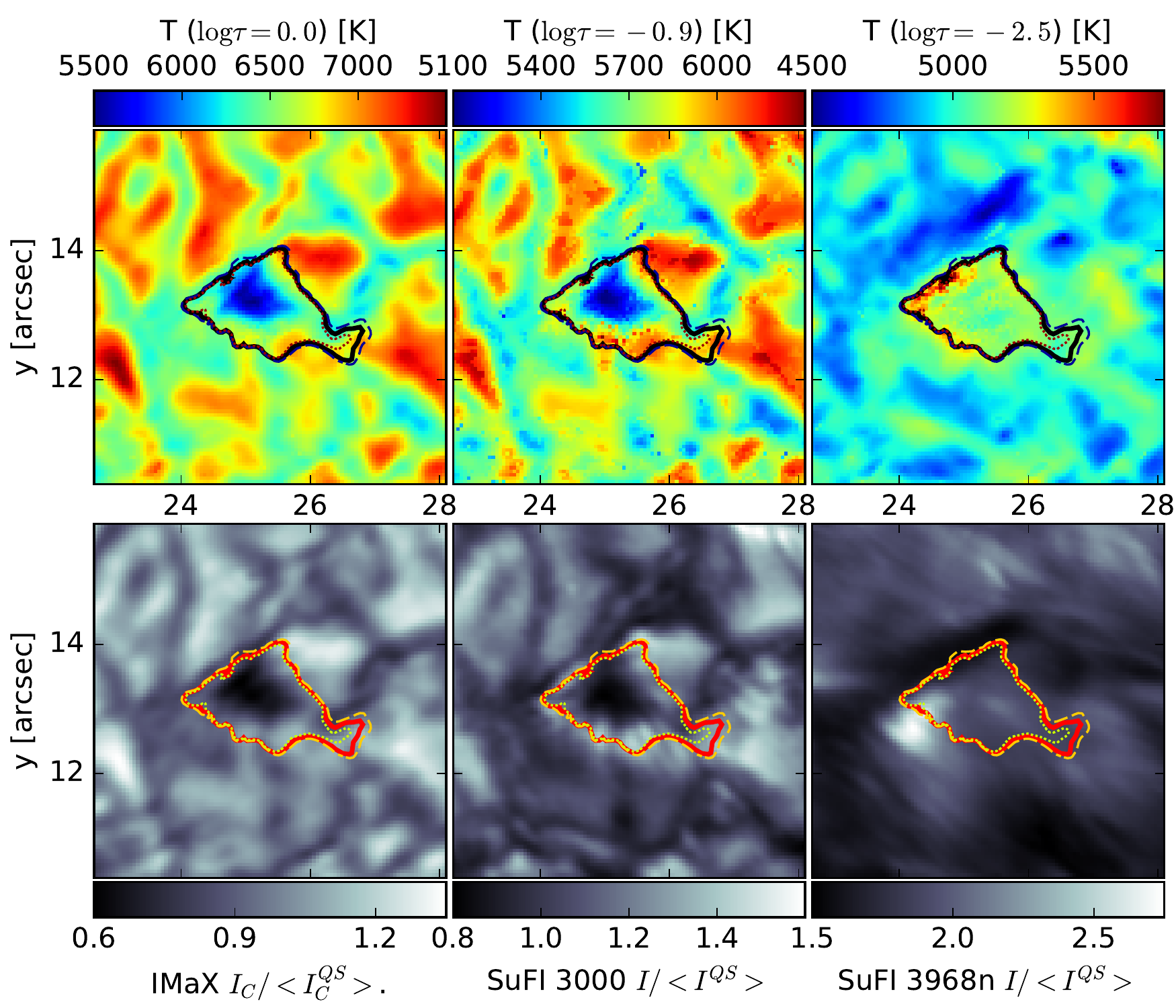}{0.8\textwidth}{}}
\caption{The same as Fig.~\ref{f8}, but for a single small pore.}
\label{f9}
\end{figure*}

Thumbnails of $T$ at $\log\tau =0, -0.9$ and $-2.5$ as well as of
IMaX continuum intensity, $I_{+227}$, and the intensity in SuFI 3000 and
3968n~\AA\ channels ("n" is for the narrow Ca~{\sc ii}~H channel introduced in
Sect.~\ref{subsec:instrum}) are plotted in Fig.~\ref{f8} for two small pores in
the top-right part of the SuFI FOV, and in Fig.~\ref{f9} for a single very
small, somewhat weaker pore in the lower part of the SuFI FOV. Overlaid on both
pores are contours of the 1400~G level of the magnetic field. In the following,
we consider this contour as an independent boundary of the pore, or more
exactly of the magnetic concentration underlying the pore.

1400~G is found to be a good compromise, on the one hand, to isolate a pore from the
bright points in its neighbourhood, while,
on the other hand, keeping as much of the pore as possible
within  the contour. E.g. if we take 1200~G or 1300~G, then the
bright points often found in the immediate vicinity of pores are included in
the contours of some of the pores. If, however, we use a larger value, such
as 1500~G or 1600~G, then we include
less of the dark part of the pore in the contour. To illustrate this we have
added also contours of 1200~G and 1600~G besides those of 1400~G in Fig.~\ref{f8}.
The 1200~G contour includes bright extensions from the
smaller (lower-left) pore in this image, while the 1600~G contour misses a
significant fraction of the upper right pore.

Interestingly,
the 1400~G contour often lies outside the actual pores as seen in, e.g.,
$I_{+227}$ or the SuFI 3000~\AA\ channel. When comparing with $I_{+227}$,
a part of this difference may have to do with the expansion of the field with
height. However, the difference in size is just as large when considering the
temperature at $\log\tau=-0.9$, which is much closer to the height at which
the magnetic field is determined.
Hence, these images demonstrate that pores are surrounded by regions of
strong magnetic field that extend well beyond  the visible boundaries of the
individual pores \citep[see also][]{MartinezPillet1997}. Of course, the
magnetic field, being determined from multiple
recordings, is more susceptible to jitter, and may not have the same high
resolution as the individual images it is compared with.
This extension of strong fields beyond the boundary of the pore still
holds even if we consider a threshold of 1500~G or even 1600~G, even as some
of the dark parts of the pores no longer lie within the 1600~G contour.

Figs.~\ref{f8} and \ref{f9} confirm that the temperature in the lower two
inversion nodes are clearly lower than in the surroundings, but are rather similar
in the upper node. Similarly, in $I_{+227}$ and at 3000~\AA\ the pores are all clearly
dark, while in Ca~{\sc ii}~H the picture is more mixed. Whereas the pore in
Fig.~\ref{f9} is clearly as bright as its surroundings and even has a particularly
bright intrusion, the pores seen in
Fig.~\ref{f8} both appear to be somewhat darker than their immediate surroundings.

We note, however, that the immediate surroundings of the pores in the Ca
images are relatively bright (see Fig.~\ref{f5}). The pores
are often surrounded by considerable magnetic flux that produces a local
brightening. In radiation coming from the lower
photosphere, some of this is visible as bright points, or particularly bright
granules or granule walls \citep[similar to faculae near the limb, e.g.,][]
{Lites2004,Keller2004,Carlsson2004}. In the low chromosphere, however,
this concentration of magnetic
flux produces enhanced brightenings in the form of mainly short fibrils
surrounding the pores. The pores are not substantially darker than other, less
bright parts of the solar surface in the Ca~{\sc ii}~H filter (see Fig.~\ref{f5}).

\begin{figure*}
\figurenum{11}
\gridline{\fig{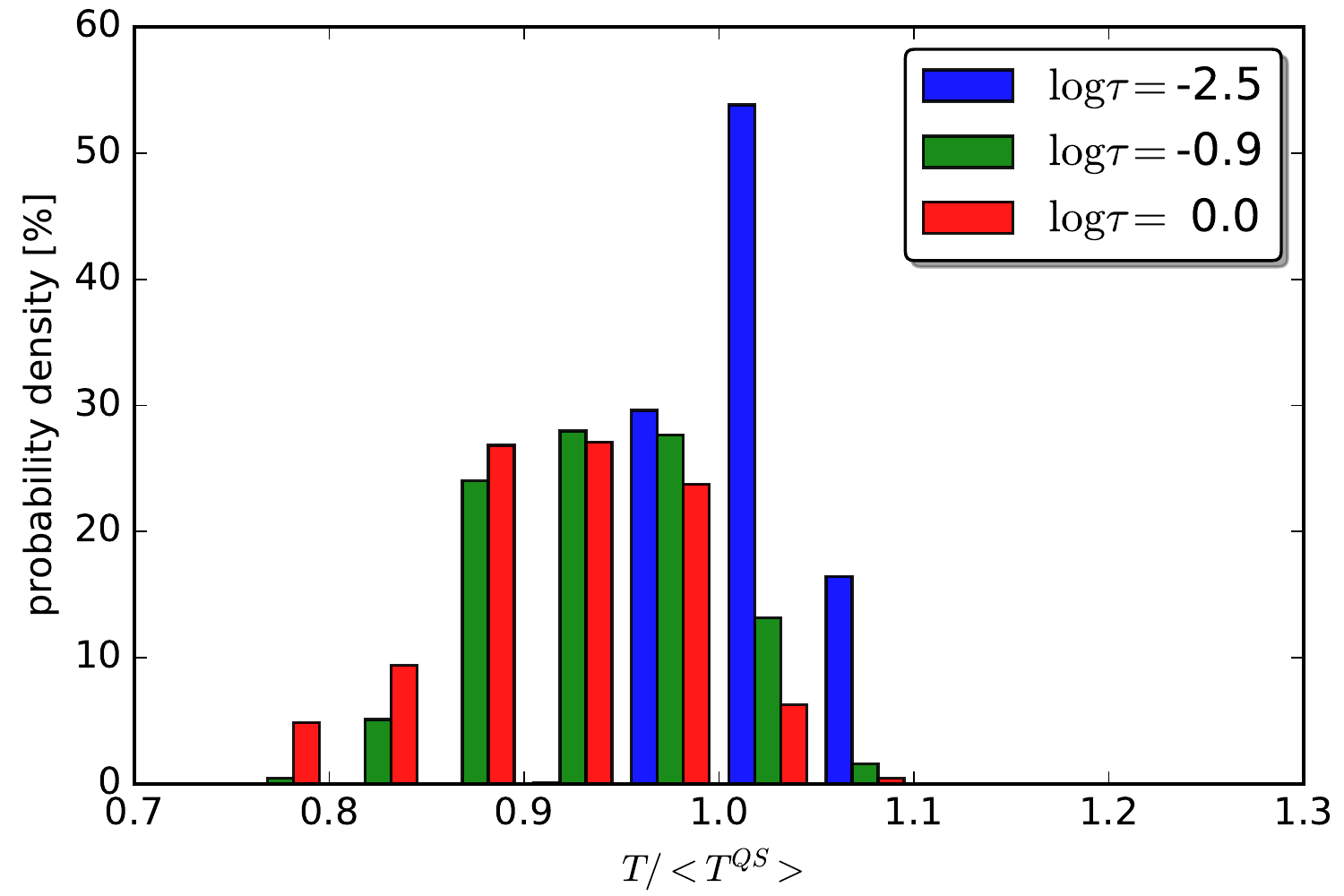}{0.49\textwidth}{(a)}
          \fig{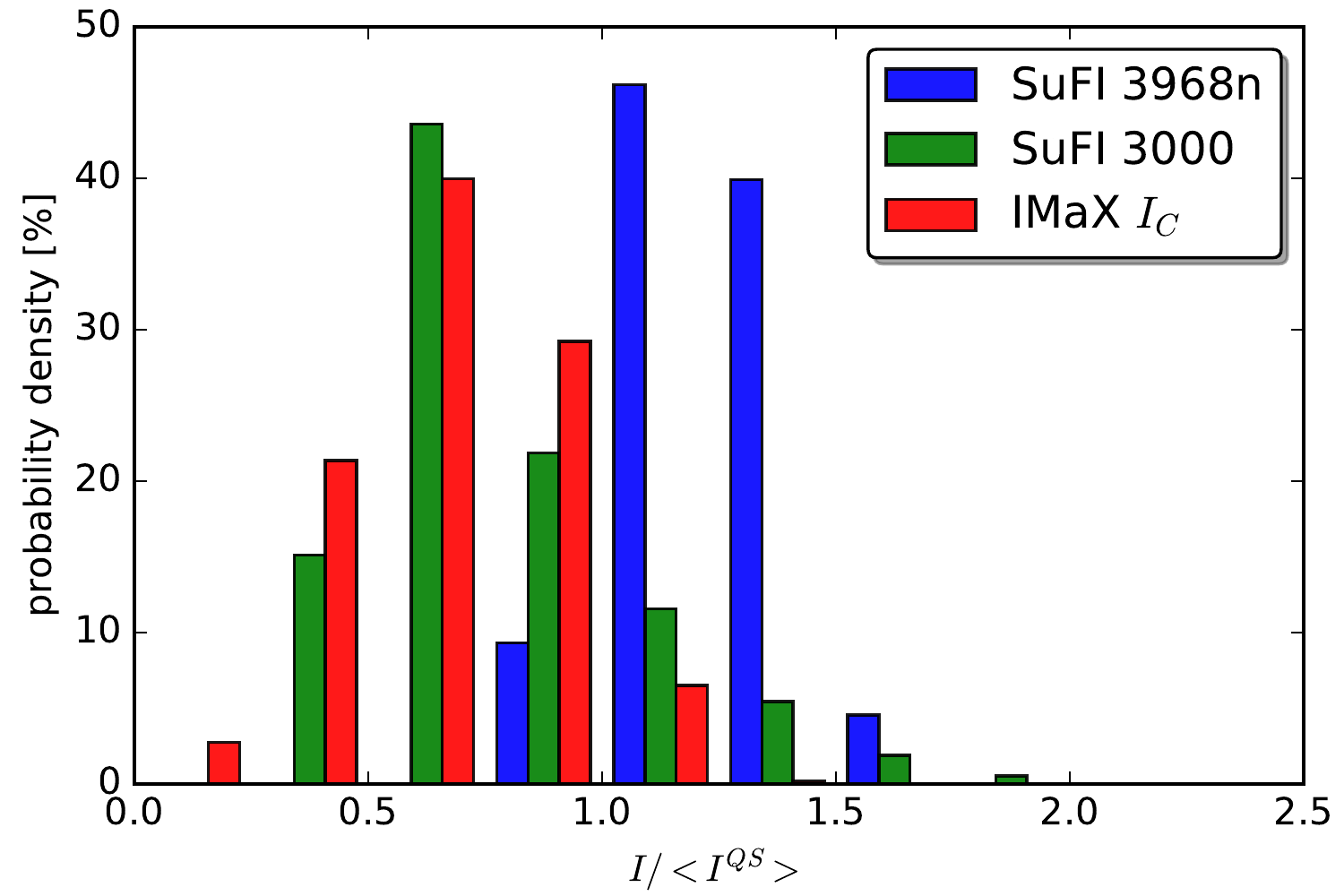}{0.49\textwidth}{(b)}
          }
\gridline{\fig{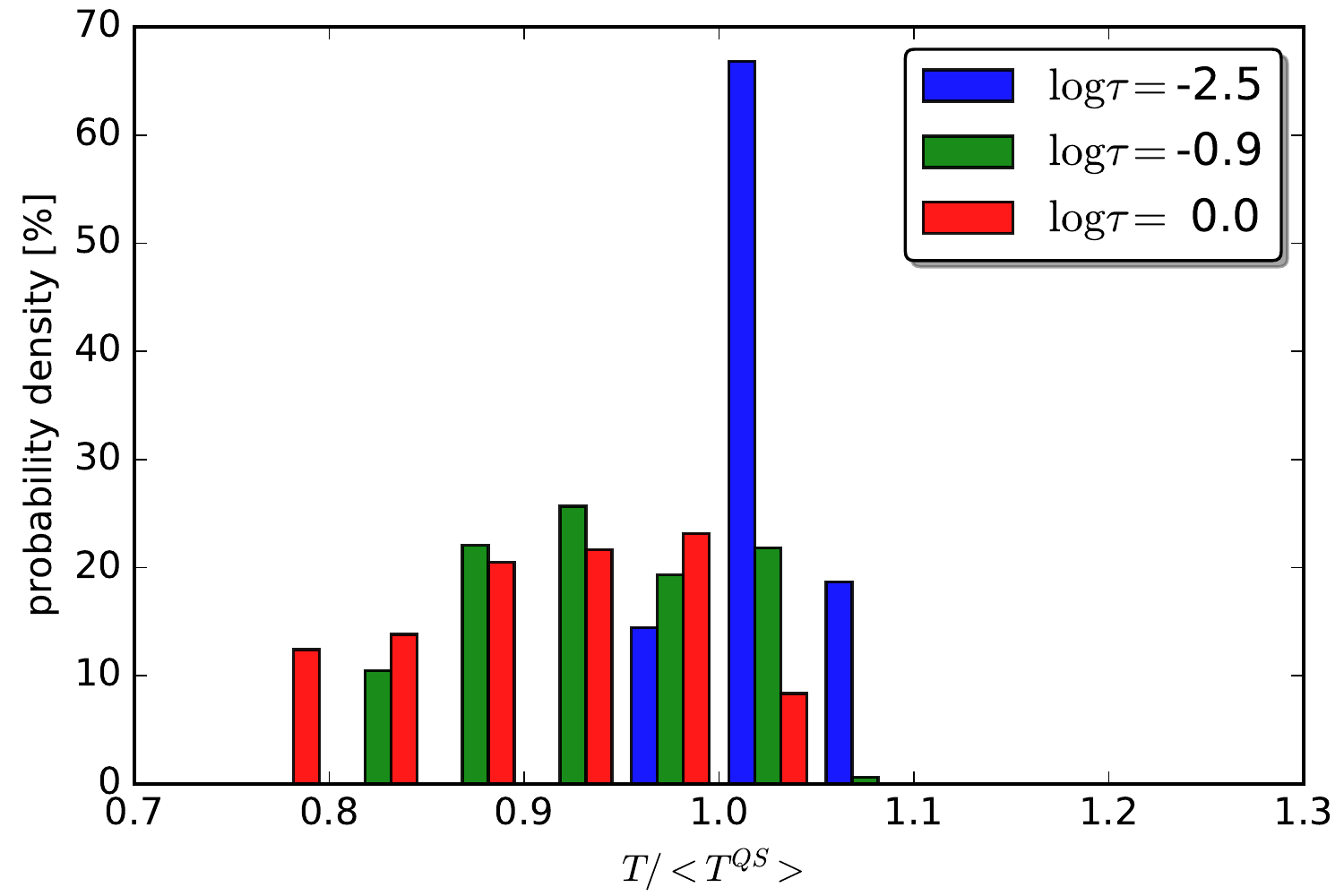}{0.49\textwidth}{(c)}
          \fig{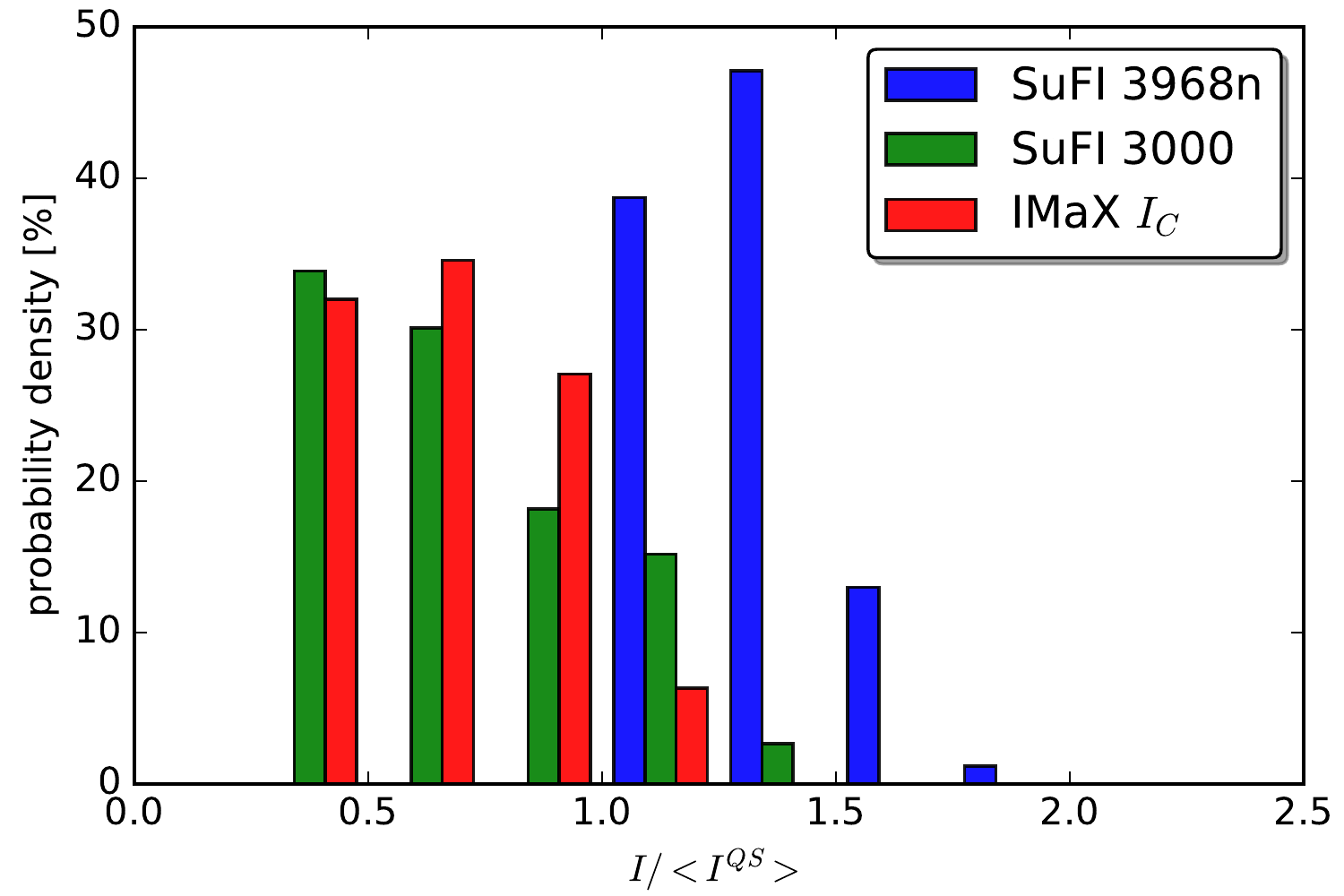}{0.49\textwidth}{(d)}
          }

\caption{(a) and (c): Histrograms of $T(\log\tau=0)$, $T(\log\tau=-0.9)$
and $T(\log\tau=-2.5$ within the
1400~G contours of the two pores displayed in Fig.~\ref{f8}.
(b) and (d): Corresponding histograms of IMaX $I_{+227}$, SuFI 3000~\AA,
SuFI 3968n~\AA. Panels (a) and (b) represent the larger pore, in the upper right
part of the images in Fig.~\ref{f8}, panels  (c) and (d) the pore in the
lower left. Histogram colours distinguish between the different $\log\tau$ layers,
or wavelengths, as labelled in the individual panels.}
\label{f10}
\end{figure*}

\begin{figure*}
\figurenum{12}
\gridline{\fig{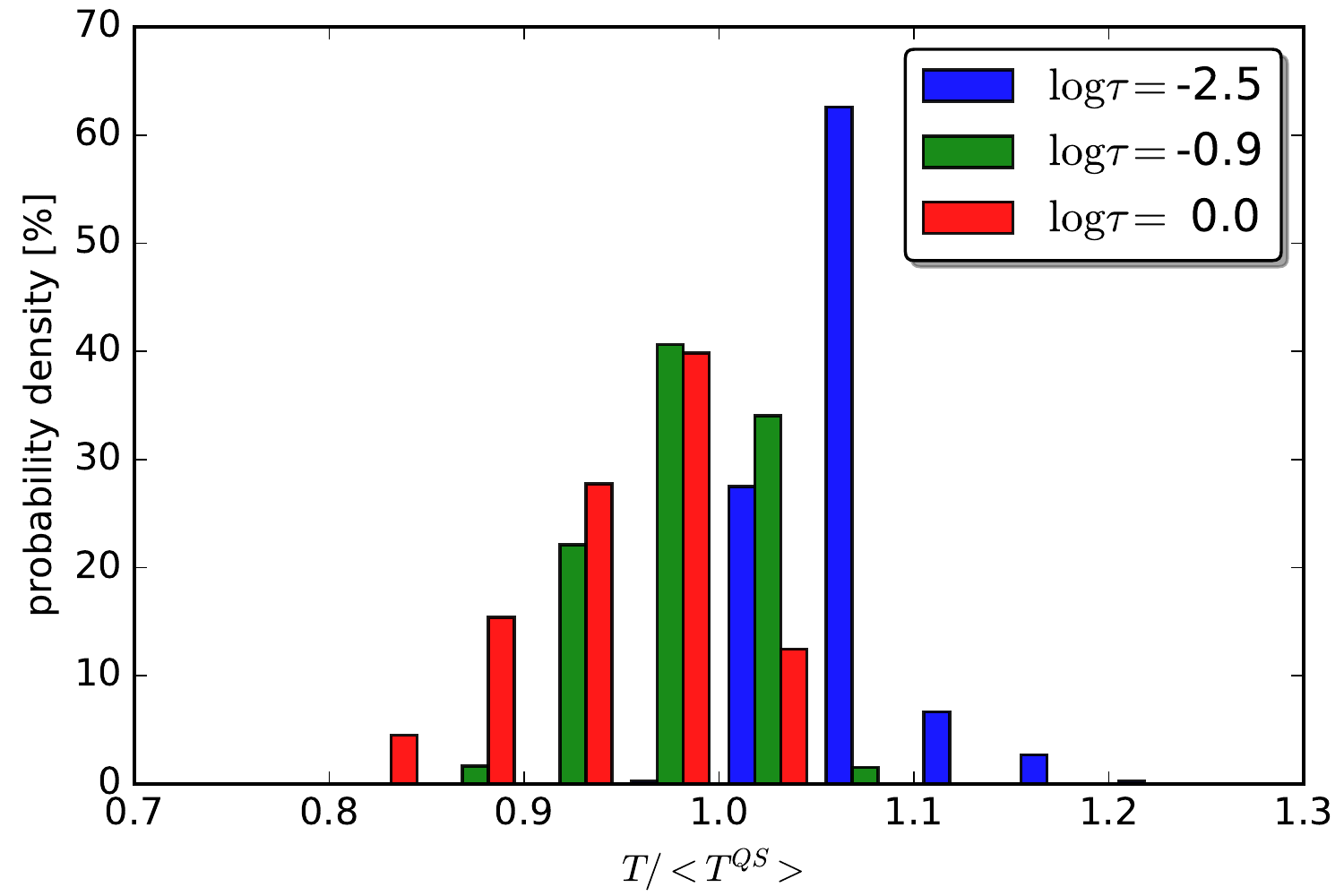}{0.49\textwidth}{(a)}
          \fig{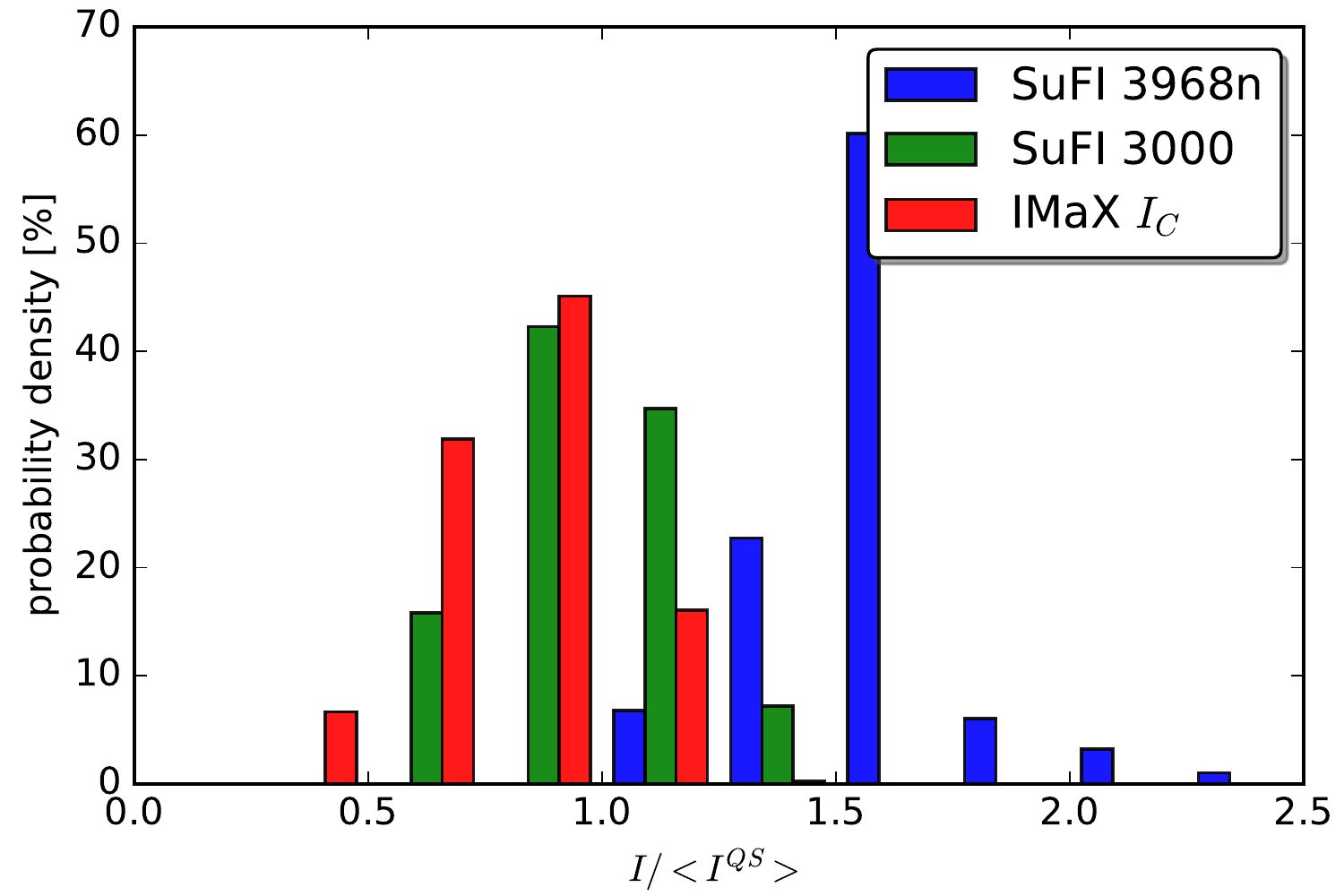}{0.49\textwidth}{(b)}
          }
\caption{The same as Fig.\ref{f10}, but for the small pore displayed in Fig.~\ref{f9}.}
\label{f11}
\end{figure*}

To study this more quantitatively, we plot the corresponding histograms in
Figs.~\ref{f10} and \ref{f11}. Each set of histograms represents either the
three temperatures, at $\log\tau=0, -0.9, -2.5$ (left panels), or the three
intensities (IMaX $I_{+227}$, SuFI 3000~\AA, SuFI 3968n~\AA; right panels) studied
here. All intensities and temperatures in Figs.~\ref{f10} and \ref{f11} are
normalized to patches of comparatively quiet Sun (i.e.\ with low amounts of
magnetic flux) near the top left and the bottom of the SuFI
FOV.

In all pores the temperatures in the two lower layers and the
intensities at the wavelengths formed deeper (IMaX $I_{+227}$ and SuFI 3000~\AA)
are predominantly lower than in the quiet Sun. The pixels in which they are
higher are often the hot walls of neighbouring granules that are seen through
a strong magnetic field. The same is true for the
intensities in the two deeper-forming wavelengths. The
temperature at the top node and the brightness in the Ca~{\sc ii}~H channel
behave quite differently, however. The temperature lies below the quiet-Sun
value for only a small fraction of the points. The same is also true for the
Ca intensity, for which the ratio of hotter and brighter points to cooler and
darker ones is even more extreme.

Hence, although the small pores partly remain visible as dark features relative
to their immediate surroundings in the  lower chromosphere, they are
approximately as bright as, or even brighter than the quiet Sun at those layers.
The most extreme enhancement is displayed by the smallest of the three pores,
with the largest being the coolest and darkest (by a small amount) of these
three.

The largest pore in the IMaX FOV (and partly in the SuFI FOV) does remain
considerably darker than the quiet Sun, even in the Ca images, which
unfortunately cover only a small part of the whole pore. Nonetheless,
the behaviour of the various pores taken together suggests
that the difference in the vertical temperature gradient within these pores
to that in the quiet Sun depends significantly on the size of the pore. The
photospheric temperature gradient is flattest for the smallest pores, with the
temperature starting nearly $1000$\,K below the average quiet-Sun value
at the solar surface and reaching the quiet-Sun value in the upper photosphere.
For large pores, however, the temperature gradient appears to be more similar to
that in the quiet Sun.

Inhomogeneities in brightness and temperature within the pores also decrease
with height. Thus, in Fig.~\ref{f8} the umbral dot-like brightenings in the
pore at the upper right of the frame
are well visible in the $I_{+227}$ at 5250.4~\AA, but only very weakly visible at
3000~\AA, which is formed only about 50~km higher, and cannot be seen at
all at 3968~\AA. This suggests that such inhomogeneities are likely of
convective origin, as proposed by \citet{Parker1979, Choudhuri1986, Schuessler2006}
for umbral dots. In sunspot umbrae a similar restriction to low heights of the thermal
enhancements/brightenings associated with umbral dots have been found by,
e.g., \cite{SocasNavarro2004}; \cite{Riethmueller2008}; \cite{Riethmueller2013a}.

\section{Conclusions and outlook} \label{sec:conclusions}

The second science flight of \sunrise{} in June 2013 allowed this
balloon-borne solar
observatory to obtain the first seeing-free observations of an active region
close to the diffraction limit of the 1\,m diameter telescope. These data are
rich in information about a variety of solar phenomena at very high spatial
resolution and at wavelengths that can partly not be accessed from the
ground. Below we list some of the findings that have been obtained from these
data, most of which are presented in various papers of this special issue.

In the present paper we have, besides providing an overview of the instrumentation
and the mission, considered thermal properties of pores as an illustration of
the capabilities of the \sunrise{}~II data. Earlier, the first ever
high-resolution images of the quiet and active Sun in the
Mg~{\sc ii}~k line, obtained during the \sunrise{}~II flight, were
reported by \cite{Riethmueller2013b} and \cite{Danilovic2014}.

In the special issue a study by \cite{Centeno2016} reports on two emerging flux
events, describing in greater detail than previously possible the interrelated
dynamics of the gas and the field, as well as the likely occurance of magnetic
reconnection during the emergence. The properties of a likely siphon flow and
of the small, initially low-lying loop connecting magnetic elements with a
pore are deduced by \cite{Requerey2016b} and the 3-D structure of the magnetic
field lines and hence of the flow vector are determined.
The properties and dynamics of moving magnetic features (MMFs) on one
side of the largest pore in the IMaX FOV are deduced by \cite{Kaithakkal2016}
and contrasted with the properties of MMFs around sunspots. The proper motion
of magnetic bright points in different parts of the quiet Sun and of an active
region are analysed by \cite{Jafarzadeh2016a}. They find very different
behaviours depending on the location, with the features moving strongly
superdiffusively in the internetwork, diffusively in the network and in between
these extremes in the active region.
\cite{Riethmueller2016} present a novel inversion technique
employing MHD simulations to provide the model atmospheres needed to compute
synthetic Stokes vectors that reproduce the observed Stokes parameters. They
illustrate the quality of the inversions by applying the technique to the
\sunrise{}~II polarimetric data.

The properties of
the slender fibrils dominating the SuFI Ca~{\sc ii}~H images are determined
by \cite{Gafeira2016a}, who show that the fibrils live much longer
than a simple analysis would suggest, if one takes into account that they often
fade away and reappear after some time. The discovery of ubiquitous transverse
waves travelling along these fibrils and carrying copious amounts of energy is
reported by \cite{Jafarzadeh2016b}, while \cite{Gafeira2016b} present the
discovery of compressible waves travelling along the fibrils, which they
identify as sausage waves. \cite{Jafarzadeh2016d} find evidence that these
slender fibrils seen in Ca~{\sc ii}~H outline a canopy of magnetic field lying
below that known from H$\alpha$ and Ca~{\sc ii} 8542~\AA\ fibril observations.

\cite{Chitta2016} observed that coronal loops are
rooted in regions with mixed-polarity fields. They provide evidence for flux
cancellation and
presence of inverse Y-shaped jets (signatures of magnetic reconnection) at the
base of coronal loops that might supply (hot) plasma to the overlying coronal
loop. They suggest a revision of the traditional picture in which each loop
footpoint is smoothly connected to unipolar regions on the solar surface.
\cite{Danilovic2016} compare an Ellerman bomb observed by \sunrise{}~II with
a similar, simulated event in which magnetic reconnection occurs at
the location of emerging flux. The 3D radiation-MHD simulation reveals the
complexity of the underlying physical process and the limitations of the
observational data. Thus, the \sunrise{}/IMaX data cannot determine the height at
which magnetic reconnection takes place. The authors also show, however, that
the velocity and magnetic
vector measured at the high resolution of \sunrise{}/IMaX reveals how shortcomings
of the MHD simulations can be overcome.

\cite{Wiegelmann2016} have computed general linear magneto-static equilibria
of the magnetic field and gas using the \sunrise{}~II/IMaX observations as a
boundary condition. In this way they obtain the magnetic field structure
in the upper atmosphere without having to assume the validity of the force-free
assumption in the solar photosphere. They computed linear magneto-static
equilibria for all the IMaX frames of the active region, without the problems
faced when modelling the magnetic field in different atmospheric layers
of the quiet Sun.


In addition to the exciting results obtained from the \sunrise{}~II flight
briefly mentioned above, the present special
issue also contains a number of papers that are based on data from the
\sunrise{}~I flight. These data are still unique in terms of consistently high
resolution polarimetric and UV time series of the quiet Sun.
Such papers include the investigation by \cite{Requerey2016a} in which the
authors uncover the tight connection between concentrated magnetic fields
and convectively driven sinks in the quiet Sun.
\cite{Kahil2016} probe the relationship
between brightness contrast at UV and visible wavelengths and the magnetic
flux in the quiet Sun, finding that the contrast keeps increasing with magnetic
flux, unlike most earlier observational results, but in qualitative agreement
with MHD simulations.
\cite{Jafarzadeh2016c} characterise the wave modes observed at
two heights in magnetic bright points, including both, compressible waves seen
in brightness fluctuations and transverse waves obtained from proper motions.
The short travel times suggest large wave speeds.
A new estimate of the flux emergence rate in the quiet Sun is obtained by
\cite{Smitha2016}. Compared with the emergence rate deduced from Hinode/SOT
data using the same technique, the emergence rate obtained from \sunrise{}~I
data is around an order of magnitude larger.

Clearly, the data from both flights of \sunrise{}
remain rich in their information content and it is expected that many
more results will be gleaned from them in the coming years. In parallel with
the scientific analysis of the data, preparations are starting for a third
flight of \sunrise{}. It is planned that on such a flight the observatory will
carry upgraded science instruments as well as new spectropolarimetric ones
that provide more information on the magnetic field and its influence on the
solar atmosphere, with particular emphasis on the solar chromosphere.

\begin{acknowledgements}
We would like to thank all current and previous team members not listed as
co-authors for their valuable contribution to the project. We thank S.~Jafarzadeh
for useful discussions and helpful comments on the manuscript. We also thank the
CSBF team for providing a perfect launch and a good recovery of the payload.
The German contribution to \sunrise{} and its reflight was funded by the
Max Planck Foundation, the Strategic Innovations Fund of the President of the
Max Planck Society (MPG), DLR, and private donations by supporting members of
the Max Planck Society, which is gratefully acknowledged. The Spanish
contribution was funded by the Ministerio de Econom\'{\i}a y Competitividad under
Projects ESP2013-47349-C6 and ESP2014-56169-C6, partially using European FEDER
funds. The National Center for Atmospheric Research is sponsored by the National
Science Foundation. The HAO contribution was partly funded through NASA grant number
NNX13AE95G. This work was partly supported by the BK21 plus program through
the National Research Foundation (NRF) funded by the Ministry of Education of
Korea. LG acknowledges research funding from the State of Lower Saxony, Germany.
SDO is a mission of NASA's Living With a Star (LWS) program. The SDO/HMI data
were provided by the Joint Science Operation Center (JSOC). The National Solar
Observatory (NSO) is operated by the Association of Universities for Research
in Astronomy (AURA) Inc. under a cooperative agreement with the National
Science Foundation.
\end{acknowledgements}


\begin{thebibliography}{}

\bibitem[Barthol et al.(2011)]{Barthol2011} Barthol, P., Gandorfer, A., Solanki, S.~K., et al. 2011,
      \solphys, 268, 1

\bibitem[Beckers and Schr\"oter(1968)]{Beckers1968} Beckers, J.~M. and Schr{\"o}ter, E.~H. 1968,
      \solphys, 4, 303

\bibitem[Bello Gonz{\'a}lez et al.(2010)]{BelloGonzalez2010} Bello Gonz{\'a}lez, N., Franz, M., Mart{\'{\i}}nez Pillet, V., et al. 2010,
      \apjl, 723, L134

\bibitem[Berkefeld et al.(2011)]{Berkefeld2011} Berkefeld, T., Schmidt, W., Soltau, D., et al. 2011,
      \solphys, 268, 103

\bibitem[Bonet et al.(2010)]{Bonet2010} Bonet, J.~A., M{\'a}rquez, I., S{\'a}nchez Almeida, J., et al. 2010,
      \apjl, 723, L139

\bibitem[Borrero et al.(2010)]{Borrero2010} Borrero, J.~M., Mart{\'{\i}}nez-Pillet, V., Schlichenmaier, R., et al. 2010,
      \apjl, 723, L144

\bibitem[Carlsson et al.(2004)]{Carlsson2004} Carlsson, M., Stein, R.~F., Nordlund, {\AA}., \& Scharmer, G.~B.\ 2004,
      \apjl, 610, L137

\bibitem[Centeno et al.(2016)]{Centeno2016}Centeno, R., Blanco Rodr\'{\i}guez, J., Del Toro Iniesta, J. C., et al. 2016,
      \apj, in press (this issue)

\bibitem[Chitta et al.(2016)]{Chitta2016}Chitta, L.~P., Peter, H., Solanki, S.~K., et al. 2016,
      \apj, in press (this issue)

\bibitem[Choudhuri(1986)]{Choudhuri1986} Choudhuri, A.~R. 1986,
      \apj, 302, 809

\bibitem[Danilovic et al.(2010)]{Danilovic2010} Danilovic, S., Beeck, B., Pietarila, A., et al. 2010,
      \apjl, 723, L149

\bibitem[Danilovic et al.(2014)]{Danilovic2014} Danilovic, S., Hirzberger, J., Riethm{\"u}ller, T.~L., et al. 2014,
      \apj, 784, 20

\bibitem[Danilovic et al.(2016)]{Danilovic2016} Danilovic, S., Solanki, S.~K., Barthol, P., et al. 2016,
      \apj, in press (this issue)

\bibitem[De Pontieu et al.(2007)]{DePontieu2007} De Pontieu, B., McIntosh, S.~W., Carlsson, M., et al. 2007,
      Science, 318, 1574

\bibitem[Frutiger(2000)]{Frutiger2000a} Frutiger, C. 2000,
      \textit{Inversion of Zeeman Split Stokes Profiles: Application to solar and stellar surface structures},
      Ph.D. Thesis, Institute of Astronomy, ETH Z\"urich, No.~13896

\bibitem[Frutiger et al.(2000)]{Frutiger2000b} Frutiger, C., Solanki, S. K., Fligge, M., \& Bruls, J. H. M. J. 2000,
      \aap, 358, 1109

\bibitem[Gafeira et al.(2016a)]{Gafeira2016a}Gafeira, R., Lagg, A., Solanki S.~K., et al. 2016a,
      \apj, submitted (this issue)

\bibitem[Gafeira et al.(2016b)]{Gafeira2016b}Gafeira, R., Jafarzadeh, S., Solanki, S.~K., et al. 2016b,
      \apj, submitted (this issue)

\bibitem[Gandorfer et al.(2011)]{Gandorfer2011} Gandorfer, A., Grauf, B., Barthol, P., et al. 2011,
      \solphys, 268, 35

\bibitem[Gonsalves \& Chidlaw(1979)]{Gonsalves1979} Gonsalves, R.~A. \& Chidlaw, R. 1979,
      \textit{Appplications of Digital Imaging Processing III, ed. A.~G. Tescher},
      Proc. SPIE, 207, 32


\bibitem[Hirzberger et al.(2010)]{Hirzberger2010} Hirzberger, J., Feller, A., Riethm{\"u}ller, T.~L., et al. 2010,
      \apjl, 723, L154

\bibitem[Hirzberger et al.(2011)]{Hirzberger2011} Hirzberger, J., Feller, A., Riethm{\"u}ller, T.~L., Gandorfer, A., \& Solanki, S.~K. 2011,
      \aap, 529, A132

\bibitem[Jafarzadeh et al.(2014)]{Jafarzadeh2014b} Jafarzadeh, S., Solanki, S.~K., Lagg, A., et al. 2014,
      \aap, 569, A105

\bibitem[Jafarzadeh et al.(2016a)]{Jafarzadeh2016a}Jafarzadeh, S., Solanki, S.~K., Cameron, R.~H., et al. 2016a,
      \apj, in press (this issue)

\bibitem[Jafarzadeh et al.(2016b)]{Jafarzadeh2016b}Jafarzadeh, S., Solanki, S.~K., Gafeira, R., et al. 2016b,
      \apj, in press (this issue)

\bibitem[Jafarzadeh et al.(2016c)]{Jafarzadeh2016c}Jafarzadeh, S., Solanki, S.~K., Stangalini, M., Cameron, R.~H., Danilovic, S. 2016c,
      \apj, submitted (this issue)

\bibitem[Jafarzadeh et al.(2016d)]{Jafarzadeh2016d}Jafarzadeh, S., Rutten, R.~J., Solanki, S.~K., et al. 2016d,
      \apj, in press (this issue)

\bibitem[Ji et al.(2012)]{Ji2012} Ji, H., Cao, W., \& Goode, P.~R.\ 2012, \apjl, 750, L25

\bibitem[Joshi et al.(2011a)]{Joshi2011a} Joshi, J., Pietarila, A., Hirzberger, J., et al. 2011a,
      \apjl, 734, L18

\bibitem[Joshi et al.(2011b)]{Joshi2011b} Joshi, J., Pietarila, A., Hirzberger, J., et al. 2011b,
      \apjl, 740, L55

\bibitem[Kahil et al.(2016)]{Kahil2016} Kahil, F., Riethm\"uller, T., Solanki, S.~K. 2016,
      \apj, submitted (this issue)

\bibitem[Kaithakkal et al.(2016)]{Kaithakkal2016}Kaithakkal, A., Riethm\"uller, T., Solanki, S.~K., et al. 2016,
      \apj, in press (this issue)

\bibitem[Katsukawa et al.(2007)]{Katsukawa2007} Katsukawa, Y., Berger, T.~E., Ichimoto, K., et al.\ 2007,
      Science, 318, 1594

\bibitem[Keller et al.(2004)]{Keller2004} Keller, C.~U., Sch{\"u}ssler, M., V{\"o}gler, A., \& Zakharov, V.\ 2004,
      \apjl, 607, L59

\bibitem[Krat et al.(1974)]{Krat1974} Krat, V.~A., Dul'Kin, L.~Z., Validov, M.~A., et al. 1974,
      Astronomicheskij Tsirkulyar, 807, 1

\bibitem[Lagg et al.(2010)]{Lagg2010} Lagg, A., Solanki, S.~K., Riethm{\"u}ller, T.~L., et al. 2010,
      \apjl, 723, L164

\bibitem[Lagg et al.(2016)]{Lagg2016} Lagg, A., Solanki, S.~K., Doerr, H.~P., et al. 2016,
      \aap, in press

\bibitem[Lites et al.(2004)]{Lites2004} Lites, B.~W., Scharmer, G.~B., Berger, T.~E., \& Title, A.~M.\ 2004,
      \solphys, 221, 65

\bibitem[Lites et al.(2008)]{Lites2008} Lites, B.~W., Kubo, M., Socas-Navarro, H., et al. 2008,
      \apj, 672, 1237-1253

\bibitem[Mart{\'{\i}}nez Gonz{\'a}lez et al.(2011)]{MartinezGonzalez2011} Mart{\'{\i}}nez Gonz{\'a}lez, M.~J., Asensio Ramos, A., Manso Sainz, R., et al. 2011,
      \apjl, 730, L37

\bibitem[Mart{\'{\i}}nez Pillet(1997)]{MartinezPillet1997} Martinez Pillet, V.\ 1997,
      \textit{1st Advances in Solar Physics Euroconference.~Advances in the Physics of Sunspots},
      Astron.\ Soc.\ Pacific Conf.\ Ser.\ Vol.~118, 212

\bibitem[Mart\'{\i}nez Pillet et al.(2011a)]{MartinezPillet2011a} Mart\'{\i}nez Pillet, V., del Toro Iniesta, J.~C., \'Alvarez-Herrero, A., Domingo, V., Bonet, J.~A., et al. 2011a,
      Sol. Phys., 268, 57

\bibitem[Mart{\'{\i}}nez Pillet et al.(2011b)]{MartinezPillet2011b} Mart{\'{\i}}nez Pillet, V., Del Toro Iniesta, J.~C., \& Quintero Noda, C. 2011b,
      \aap, 530, A111

\bibitem[Mehltretter(1974)]{Mehltretter1974} Mehltretter, J.~P.\ 1974,
      \solphys, 38, 43

\bibitem[Orozco Su{\'a}rez et al.(2007)]{Orozco2007} Orozco Su{\'a}rez, D., Bellot Rubio, L.~R., del Toro Iniesta, J.~C., et al.\ 2007,
      \apjl, 670, L61

\bibitem[Parker(1979)]{Parker1979} Parker, E.~N.\ 1979,
      \apj, 234, 333

\bibitem[Paxman et al.(1992)]{Paxman1992} Paxman, R.~G., Schulz, T.~J., \& Fienup, J.~R. 1992,
      J.\ Opt.\ Soc.\ Am., 9, 1072

\bibitem[Paxman et al.(1996)]{Paxman1996} Paxman, R.~G., Seldin, J.~J., Loefdahl, M.~G., Scharmer, G.~B., \& Keller, C.~U. 1996,
      \apj, 466, 1087

\bibitem[Pietarila et al.(2009)]{Pietarila2009} Pietarila, A., Hirzberger, J., Zakharov, V., \& Solanki, S.~K.\ 2009,
      \aap, 502, 647

\bibitem[Quintero Noda et al.(2016)]{QuinteroNoda2016} Quintero Noda, C., Shimizu, T., Ruiz Cobo, B., et al.\ 2016,
      \mnras, 460, 1476

\bibitem[Rayleigh(1879)]{Rayleigh1879} Rayleigh, Lord, F.~R.~S. 1879,
      \textit{Investigations in optics, with special reference to the spectroscope},
      Phil. Mag., 8, 261-274

\bibitem[Requerey et al.(2014)]{Requerey2014} Requerey, I.~S., Del Toro Iniesta, J.~C., Bellot Rubio, L.~R., et al.\ 2014,
      \apj, 789, 6

\bibitem[Requerey et al.(2016a)]{Requerey2016a}Requerey, I., Del Toro Iniesta, J.~C., Bellot Rubio, L.~R., et al. 2016a,
      \apj, in press (this issue)

\bibitem[Requerey et al.(2016b)]{Requerey2016b}Requerey, I., Ruiz Cobo, B., Del Toro Iniesta, J.~C., et al. 2016b,
      \apj, in press (this issue)

\bibitem[Riethm{\"u}ller et al.(2008)]{Riethmueller2008} Riethm{\"u}ller, T.~L., Solanki, S.~K., \& Lagg, A.\ 2008,
      \apjl, 678, L157

\bibitem[Riethm{\"u}ller et al.(2010)]{Riethmueller2010} Riethm{\"u}ller, T.~L., Solanki, S.~K., Mart{\'{\i}}nez Pillet, V., et al.\ 2010,
      \apjl, 723, L169

\bibitem[Riethm{\"u}ller et al.(2013a)]{Riethmueller2013a} Riethm{\"u}ller, T.~L., Solanki, S.~K., van Noort, M., \& Tiwari, S.~K. 2013a,
      \aap, 554, A53

\bibitem[Riethm{\"u}ller et al.(2013b)]{Riethmueller2013b} Riethm{\"u}ller, T.~L., Solanki, S.~K., Hirzberger, J., et al.\ 2013b,
      \apjl, 776, L13

\bibitem[Rieth\-m\"ul\-ler et al.(2014)]{Riethmueller2014} Rieth\-m\"ul\-ler, T.~L., Solanki, S.~K., Berdyugina, S.~V., Sch\"ussler, M., Mart\'{\i}nez Pillet, V., et al. 2014,
      \aap, 568, A13

\bibitem[Riethm\"uller et al.(2016)]{Riethmueller2016} Riethm\"uller, T.~L., Solanki, S.~K., Barthol, P., et al. 2016,
      \apj, in press  (this issue)

\bibitem[Scharmer et al.(2002)]{Scharmer2002} Scharmer, G.~B., Gudiksen, B.~V., Kiselman, D., L{\"o}fdahl, M.~G., \& Rouppe van der Voort, L.~H.~M.\ 2002,
      \nat, 420, 151

\bibitem[Scharmer et al.(2011)]{Scharmer2011} Scharmer, G.~B., Henriques, V.~M.~J., Kiselman, D., \& de la Cruz Rodr{\'{\i}}guez, J.\ 2011,
      Science, 333, 316

\bibitem[Sch{\"u}ssler \&\ V\"ogler(2006)]{Schuessler2006} Sch{\"u}ssler, M., V\"ogler, A.\ 2006,
      \apjl, 641, L73

\bibitem[Smitha et al.(2016)]{Smitha2016}Smitha, H.~N., Anusha, L.~S., Solanki, S.~K., Riethm\"uller, T.~L. 2016,
      \apj, in press  (this issue)

\bibitem[Socas-Navarro et al.(2004)]{SocasNavarro2004}
      Socas-Navarro, H., Mart{\'{\i}}nez Pillet, V., Sobotka, M., \& V{\'a}zquez, M.\
      2004, \apj, 614, 448

\bibitem[Solanki(1987)]{Solanki1987} Solanki, S. K. 1987,
      \textit{The Photospheric Layers of Solar Magnetic Fluxtubes},
      Ph.D. Thesis, Institute of Astronomy, ETH Z\"urich, No.~8309

\bibitem[Solanki et al.(2010)]{Solanki2010} Solanki, S.~K., Barthol, P., Danilovic, S., et al.\ 2010,
      \apjl, 723, L127

\bibitem[Steiner et al.(2010)]{Steiner2010} Steiner, O., Franz, M., Bello Gonz{\'a}lez, N., et al.\ 2010,
      \apjl, 723, L180

\bibitem[Tsuneta et al.(2008)]{Tsuneta2008} Tsuneta, S., Ichimoto, K., Katsukawa, Y., et al.\ 2008,
      \solphys, 249, 167

\bibitem[van Noort et al.(2005)]{vanNoort2005} van~Noort, M., Rouppe van der Voort, L., \& L\"ofdahl, M.~G. 2005,
      \solphys, 228, 191

\bibitem[van Noort et al.(2013)]{vanNoort2013} van Noort, M., Lagg, A., Tiwari, S.~K., \& Solanki, S.~K.\ 2013,
      \aap, 557, A24

\bibitem[Wiegelmann et al.(2010)]{Wiegelmann2010} Wiegelmann, T., Solanki, S.~K., Borrero, J.~M., et al.\ 2010,
      \apjl, 723, L185

\bibitem[Wiegelmann et al.(2016)]{Wiegelmann2016}Wiegelmann, T., Neukirch, T., Nickeler, D.~H., et al. 2016,
      \apj, submitted (this issue)

\bibitem[Yelles Chaouche et al.(2011)]{Yelles2011} Yelles Chaouche, L., Moreno-Insertis, F., Mart{\'{\i}}nez Pillet, V., et al.\ 2011,
      \apjl, 727, L30


\end{thebibliography}
\end{document}